\documentclass{elsart}
\usepackage{graphicx}
\usepackage{amssymb}
\usepackage{amsmath}
\begin{document}
\begin{frontmatter}
\journal{Physica D}
\title{Time-frequency analysis of chaotic systems}
\date{\today}
\author{C. Chandre$^1$, S. Wiggins$^2$, T. Uzer$^1$}
\address{$^1$Center for Nonlinear Science, School of Physics,
Georgia Institute of Technology, Atlanta, Georgia 30332-0430, U.S.A.}
\address{$^2$Department of Mathematics, University of Bristol, Bristol BS8
1TW, U.K.}
\begin{abstract}        
We describe a method for analyzing the phase space structures 
of Hamiltonian systems. 
This method is based on a time-frequency decomposition of a trajectory using 
wavelets. The ridges of the time-frequency landscape of a 
trajectory, also called instantaneous frequencies, enable us to analyze the 
phase space structures. In particular, this method detects resonance trappings 
and  transitions and allows a characterization of the notion of weak and strong 
chaos. We illustrate the method with the trajectories of the standard map and 
the hydrogen atom in crossed magnetic and elliptically polarized 
microwave fields. \end{abstract} 
\begin{keyword}
Time-frequency analysis \sep wavelets \sep Hamiltonian systems
\PACS 05.45.-a \sep  32.80.Rm 
\end{keyword}
\end{frontmatter}

\section{Introduction}

The phase space of a typical Hamiltonian system is a mixture of periodic, 
quasiperiodic and chaotic trajectories, and the stable and unstable manifolds 
of these trajectories. The analysis of the trajectories of those systems is 
central to many branches in physics, ranging from celestial mechanics to 
chemistry. Several methods have been designed to discriminate between regular 
(periodic or quasiperiodic) and chaotic trajectories. For two degrees of 
freedom, the most commonly used method is a Poincar\'e section which is a plane 
section of the three dimensional energy surfaces. These sections give pictures 
of phase space and allow to analyze qualitatively the stability properties of 
the system by revealing, e.g., the invariant tori, the stable islands 
surrounding elliptic periodic orbits and the spatial extent of chaotic 
trajectories. However, these sections do not provide very accurate information 
and the visual inspection of these sections for higher dimensional systems  is 
less straightforward since, e.g., for Hamiltonian systems with three degrees of 
freedom, these sections are four dimensional.\\ Much of our understanding of 
nonlinear systems is based on their frequencies and  especially resonances 
between them (perturbation theorems, intramolecular 
dynamics~\cite{mart87,milc98}, Laskar's work in celestial 
mechanics~\cite{lask90,lask99,corr01} and particle accelerators~\cite{robi00}). 
One way to obtain accurate information on the stability of the system is to 
associate with a given trajectory an indicator such as a frequency (in general, 
several indicators are associated with a given trajectory) and construct, e.g., 
a frequency representation of phase space. In this frequency space, regular 
regions are characterized by a very small variation of the frequency as the 
time interval on which this frequency is computed increases (diffusion constant 
in Laskar's work~\cite{lask93}). The main advantage of this method is that it is 
generalizable to systems with any number of degrees of freedom~\cite{lask90}.\\ 
However, a frequency representation of the phase space does not provide the 
history and the local properties of a trajectory. The computation of asymptotic 
quantities (such as Lyapunov exponents, entropy or fractal dimensions) does not 
provide dynamical information. These quantities only reflect the asymptotic 
behavior of the system.  For many purposes infinite-time quantities are not 
sufficient : rather than knowing whether or not a trajectory is chaotic, it is 
naturally desirable to say when, where, and to what degree a trajectory is 
chaotic. \\ In this article, we describe a method that gives a {\em 
time-frequency} representation of trajectories, i.e., that provides dynamical 
information on the system.  The idea is to follow a given trajectory and see if 
the phase space structures where the trajectory passes by are revealed by 
analyzing a single trajectory. For quasiperiodic and chaotic systems, 
time-frequency representation is revealing because it shows when and how motion 
can be trapped in some resonance zones associated with quasiperiodicity. It also 
show the degree of chaoticity of some part of phase space visited by the 
trajectory. \\

In a nutshell, we associate a set of
time-varying frequencies with the trajectory by decomposing it on
a set of elementary functions, the wavelets, which are localized in
time and frequency. Wavelet decomposition has been used to analyze
trajectories of Hamiltonian systems in celestial mechanics~\cite{mich96}
and in molecular dynamics~\cite{aska96,vela01}. In Refs.~\cite{mich96,vela01},
the main
frequency was extracted by computing the frequency curve where the
modulus of the wavelet transform is maximum. However, stopping at
this main frequency can give misleading or wrong information
about resonance transitions. Instead, we have found that the
relevant information of the trajectory is contained in the
\textit{ridges} of the landscape formed by the magnitude of the
coefficients of the time-frequency decomposition~\cite{delp92}. A
ridge curve (which has a certain time length) corresponds to
time-frequency energy localization and is also called
\textit{instantaneous frequency} curve. We show that the instantaneous 
frequencies point to resonance trappings and transitions, and give a clear 
characterization of the degree of chaoticity.\\

In Sec.~\ref{sec2}, we describe the time-frequency method and the extraction of 
ridges by two methods: the windowed Fourier transform and the continuous 
wavelet transform. In Sec.~\ref{sec3}, we present briefly the two models which 
are used to illustrate the method in  Sec.~\ref{sec4}: the first one is the 
standard map which is used as a benchmark model for the method since the phase 
space structures of this model are well-known, and the second one is the 
hydrogen atom driven by a elliptically polarized microwave field which is a 
basic and realistic problem in atomic physics with three degrees of freedom.\\ 
We show how time-frequency analysis reveals the phase space structures: For two 
degrees of freedom, we compare this analysis with the Poincar\'e surface of 
section, and we show that the time-frequency analysis can be carried out for 
higher dimensional systems.\\
In Appendix 
A, we give an approximation of the ridge curves for a quasiperiodic trajectory 
and in Appendix B, we compute the effect of finite time on the computation of 
the ridges.  

\section{Time-frequency analysis}
\label{sec2}

Given a real signal $f(t)$ which can be, e.g., one coordinate of
the system (one position or one momentum for instance), two methods are used to 
compute a time-frequency (denoted $u$ - $\xi$) representation of $f$: windowed 
Fourier transform and continuous wavelet transform.  We assume that we have 
computed the trajectory for all times $t\in \mathbb{R}$. The effect of finite 
time interval is analyzed in Appendix B.  For more details on time-frequency 
analysis, we refer to Refs.~\cite{Bmall99,Bcarm98,delp92,hess96,huan98}

\subsection{Windowed Fourier transform} The windowed Fourier
transform (also called Gabor transform) of $f(t)$ is given by
\begin{equation} \label{eqn:wft} S
f(u,\xi)=\int_{-\infty}^{+\infty} f(t) g(t-u)e^{-i\xi t} dt ,
\end{equation}
where the window $g$ is chosen to be a Gaussian window $g(t)=
e^{-t^2/2\sigma^2}/(\sigma^2 \pi)^{1/4}$. We notice that in
Laskar's work~\cite{lask90,lask99}, the Hanning filter
$\chi(t)=1-\cos(2\pi t/T)$ over a time interval $[0,T]$, is
chosen rather than a Gaussian filter. The Hanning filter has the
advantage of a finite support whereas the Gaussian filter
has a minimal time-frequency resolution. We will consider an
energy density in the time-frequency plane called spectrogram
$$ P_S f(u,\xi)=\vert S f(u,\xi)\vert^2.$$
We notice that $P_S f(u,-\xi)=P_S f(u,\xi)$ for $f$ real. It is thus sufficient 
to consider the positive frequency part of the time-frequency plane. 
\subsubsection{The algorithm} The algorithm to compute $Sf(u,\xi)$ uses one fast 
Fourier transform (FFT) of $f(\cdot +u)g(\cdot)$ at each time $u$ since $$
Sf(u,\xi)=e^{-i\xi u}\int_{-\infty}^{+\infty} f(t+u)g(t)e^{-i\xi t} dt.
$$
\subsubsection{Time-frequency resolution}
The resolution of the transform is constant in time and in frequency.
The time spread around a point $(u,\xi)$ in the time-frequency plane, defined as 
$$\sigma_{time}(u,\xi)=\left(\int_{-\infty}^{+\infty} (t-u)^2 
|g_{u,\xi}(t)|^2 dt\right)^{1/2},$$
where $g_{u,\xi}(t)=g(t-u)e^{-i\xi t}$, is independent of $u$ and $\xi$. For a 
Gaussian window $g(t)=e^{-t^2/2\sigma^2}/(\sigma^2 \pi)^{1/4}$, it is 
equal to $\sigma/\sqrt{2}$. The frequency spread around $(u,\xi)$ defined as 
$$ \sigma_{freq}(u,\xi)=\left( \frac{1}{2\pi} \int_{-\infty}^{+\infty} 
(\omega- \xi)^2 |\hat{g}_{u,\xi}(\omega)|^2 d\omega\right)^{1/2} $$ 
does not depend on $u$ and $\xi$ since 
$\hat{g}_{u,\xi}(\omega)=\hat{g}(\omega-\xi)e^{iu(\xi-\omega)}$. For a Gaussian 
window,  it is equal to $1/(\sigma\sqrt{2})$. We notice that the 
product of the time spread with the frequency spread at a given point $(u,\xi)$ 
in the time frequency-plane is constant and larger than 1/2 
(Heisenberg uncertainty). It is minimum for a Gaussian window. 

\subsection{Continuous wavelet transform}

The continuous wavelet transform of $f(t)$ gives a time-scale
representation of the trajectory and is given by
\begin{equation}
\label{eqn:cwt} W
f(u,s)=\frac{1}{\sqrt{s}}\int_{-\infty}^{+\infty} f(t) \psi^*
\left( \frac{t-u}{s}\right) dt ,
\end{equation}
where the mother wavelet $\psi$ is chosen to be a Gabor (modulated
Gaussian) wavelet, also called Morlet-Grossman wavelet: $\psi(t)=
e^{i \eta t}e^{-t^2/2\sigma^2}/(\sigma^2 \pi)^{1/4}$. The
transform depends on $\eta$ which is the center frequency of the
wavelet
$$
\eta=\frac{1}{2\pi} \int_{-\infty}^{+\infty} \omega \vert 
\hat{\psi}_{u,s}(\omega)\vert ^2 d\omega,
$$
where $\psi_{u,s}(t)=s^{-1/2}\psi[(t-u)/s]$.
The time-frequency representation is obtained by the
relation between the scale $s$ and the frequency $\xi$:
\begin{equation}
\xi=\frac{\eta}{s}.
\end{equation} 
We will consider the normalized scalogram
$$ P_W f(u,\xi=\eta/s)=\frac{1}{s}\vert W
f(u,s)\vert^2,$$ which can be interpreted as the energy density in
the time-frequency plane.   Again, we notice that $P_W f(u,-\xi)=P_W f(u,\xi)$. 
For the computation of the wavelet decomposition, we only consider the positive 
frequency part of the time-frequency plane.                
\subsubsection{ The 
wavelet algorithm} In order to use a fast wavelet algorithm, we consider that 
$f$ and $\psi$ are periodic (and hence $Wf(u,s)$ is periodic in the variable 
$u$) with period $T$, the time span of the trajectory. By taking the Fourier 
transform of  Eq.~(\ref{eqn:cwt}), one has: $$ \hat{Wf}(\omega,s)=\sqrt{s} 
\hat{f}(\omega) \hat{\psi}^*(s\omega), $$ Since $\hat{\psi}(\omega)$ is known 
exactly, one has to compute $\hat{f}$ by FFT and then retrieve $Wf$ by an 
inverse FFT. For this computation, the scales are chosen according to a dyadic 
sampling $s_{i,j}= 2^{i+j/n}T/N$ for $i=1,\ldots,m$ and $j=1,\ldots,n$ where $m$ 
is the number of octaves, $n$ is the number of voices per octave and $N$ is the 
number of points in the time series on the interval $[0,T]$. The relative error 
produced by the algorithm on the computation of the frequency is then $\Delta 
\omega /\omega =1-2^{-1/n}\approx (\log 2) /n$. This error decreases when the 
number of voices per octave is increased. We chose $n\approx 300$ is our 
computations (which gives a relative error of $2 \cdot 10^{-3}$).  In what 
follows, we choose $m=\log_2 N -3$ where $N$ is the total number of points in 
the series. 

In order to compute the continuous wavelet transform, we use the routines of 
\textsc{WaveLab}, written in \textsc{Matlab}~\cite{wavelab}. 

{\em Remark:} In practice we choose $\eta=2$ and $\sigma=6$. We notice 
that the wavelet does not satisfy the admissibility condition $\hat{\psi}(0)=0$ 
which is required for reconstructing the trajectory. However   
$\hat{\psi}(0)=(4\pi \sigma^2)^{1/4} e^{-\sigma^2\eta^2/2}$ is negligible since 
$\sigma \eta $ is large enough  (e.g, if $\sigma\eta=12$, 
$\vert \hat{\psi}(0)\vert <10^{-30}$).  If one wants to satisfy the 
admissibility condition, the wavelet has to be chosen according to
$$
\psi(t)=c e^{-t^2/(2\sigma^2)}(e^{i\eta t} -e^{-\sigma^2\eta^2/2}),
$$
with $c=(\pi\sigma^2)^{-1/4} 
(1-2e^{-3\sigma^2\eta^2/4}+e^{-\sigma^2\eta^2})^{-1/2}$. The relation between 
the scale and the frequency is changed. It becomes
$\xi=(1+e^{-\sigma^2\eta^2})\eta/s$ up to the first order. For the parameters 
we choose, this error in determining the frequency is negligible. 

\subsubsection{Time-frequency resolution}
The time-frequency resolution depends on the scale of the wavelet. 
The time spread around a point $(u,s)$ in the time-scale plane, defined as 
$$\sigma_{time}(u,s)=\left(\int_{-\infty}^{+\infty} (t-u)^2 
|\psi_{u,s}(t)|^2 dt\right)^{1/2},$$
is independent of $u$ (the moment 
of observation of the frequency content) and is proportional to the scale 
$s=\eta/\xi$ for a mother wavelet of the form $\psi(t)=e^{i\eta t} g(t)$. 
For a Gabor wavelet, $\sigma_{time}$ is equal to $\sigma\eta/(\sqrt{2}\xi)$. 
From this equation, we see that for low frequency the time spread of the wavelet 
is large, whereas for high frequencies, it is small. This allows a more accurate 
computation of the frequency since for low frequencies a larger time interval is 
needed (large enough to contain several oscillations at that frequency). 
Increasing the parameters $\sigma$ and $\eta$ increases the time spread which 
varies as $1/\xi$. The frequency spread around $\xi=\eta/s$ defined as $$ 
\sigma_{freq}(u,s)=\left( \frac{1}{2\pi} \int_{-\infty}^{+\infty} (\omega- 
\eta/s)^2 |\hat{\psi}_{u,s}(\omega)|^2 d\omega\right)^{1/2}, $$ is 
independent of $u$ and is linearly dependent on the 
frequency for a mother wavelet of the form $\psi(t)=e^{i\eta t} g(t)$. For a 
Gabor wavelet, $\sigma_{freq}$  is equal to $\xi/(\sigma\eta \sqrt{2})$. 
We notice that the product of the time spread with the frequency spread at a 
given time-frequency point is independent of $u$ and $\xi=\eta/s$ and is 
larger than $1/2$ (Heisenberg uncertainty). The Heisenberg 
uncertainty is minimum for the Gabor wavelet. Increasing or decreasing the 
product of the parameters $\sigma \eta$ is a compromise between the time 
resolution and the frequency resolution.

\subsection{Differences between windowed Fourier transform and
wavelet transform}            
Both methods have time-frequency resolution limitations for the
determination of the instantaneous frequencies. The windowed
Fourier transform relies on an a priori choice of length of the
window $\sigma$.  Any event (trapping, transition, etc.) happening
on short time-scales (less than $\sigma$) or with small
frequencies (less than $\sigma^{-1}$) is missed by this method. On
the contrary, the main advantage of the wavelet method  is that it follows
the rapid variations of the instantaneous frequencies since it
adapts the length of the window according to the
frequency~\cite{Bmall99}.  \\
For quasiperiodic trajectories, there are practically no differences between 
both methods.  However, for chaotic trajectories where variations of 
frequency with respect to time are expected, the wavelet basis is more adapted 
since it leads to a better time-frequency resolution. In what follows, we choose 
the wavelet basis in order to compute the time-frequency content of the 
trajectories. 

\subsection{Ridge plots}
\label{sec24}
 The different main frequencies of the trajectory can be obtained
by looking at the \textit{ridges} of the spectrogram or normalized
scalogram, also called \textit{instantaneous frequencies}. These
ridges are local maxima with respect to the frequency $\xi$, of
the energy density in the time-frequency plane which is the spectrogram or the 
normalized scalogram. \\                             For a periodic trajectory 
$f(t)=e^{i\omega t}$, the spectrogram $$ P_S f(u,\xi)=2\sigma 
\sqrt{\pi}e^{-\sigma^2(\omega-\xi)^2}$$ and the normalized scalogram $$P_W
f(u,\xi)=2\sigma\sqrt{\pi} e^{-\sigma^2\eta^2(\omega/\xi -1)^2}$$
are maximum for $\xi=\omega$, independently of the time $u$. Therefore, the 
ridge plot will present a single flat ridge located at $\xi=\omega$. \\
For a quasi-periodic trajectory $f(t)=\sum_k A_k e^{i\omega_k t}$, the
spectrogram and the normalized scalogram present a sum of
localized peaks and interference terms~: The wavelet transform of
$f$ is given by the sum of the wavelet transform of $A_k
e^{i\omega_k t}$ since the wavelet transform is linear. Therefore,
the normalized scalogram is a sum of peaks located at
$\xi=\omega_k$ and of width proportional to
$\omega_k/(\eta\sigma)$, with some additional interference terms
of the form $$ I_{kl}=A_l A_k^* e^{-\sigma^2\eta^2
(\omega_l/\xi-1)^2/2}e^{-\sigma^2\eta^2
(\omega_k/\xi-1)^2/2}e^{i(\omega_l-\omega_k) u}.$$ The effect of
these terms is visible if $(a)$ at least two amplitudes $A_k$ and
$A_l$ are large enough, and $(b)$ if the difference of their
frequencies $\omega_l-\omega_k$ is smaller than
$\omega_l/(\eta\sigma)$ (see Appendix A). In order to distinguish two frequency
components $\omega_l$ and $\omega_k$ of amplitude $A_k$ and $A_l$
of the same order, the parameters of the wavelet must be chosen
according to the condition $\eta\sigma \geq
(\omega_l+\omega_k)/|\omega_l -\omega_k|$.  
 The time-frequency landscape has
\textit{approximately} a set of flat ridges (if the
interferences are negligible) at $\xi=\omega_k$. \\
{\em Remark: Laskar's frequency map analysis.} In order to determine accurately 
the frequencies of a quasiperiodic trajectory and the location of the ridges, we 
use Laskar's frequency map analysis~\cite{lask90,lask93,lask99} which is a 
Fourier based analysis of quasiperiodic functions $f(t)=\sum_k a_k e^{i\omega_k 
t} $ known on a finite interval $[0,T]$. The approximation of the main frequency 
$\omega_1$, called $\omega'_1$ is obtained by maximizing the modulus of $$ 
\phi(\omega)=\langle f , e^{i\omega t} \rangle = \frac{1}{T}\int_0^T f(t) \chi 
(t) e^{-i\omega t}, $$ where $\chi(t)=1-\cos(2\pi t/T)$ is a Hanning filter. The 
amplitude $a'_1$ of the component with frequency $\omega'_1$ is obtained by 
projection on $e^{i\omega'_1t}$. Laskar proved that this method gives 
the main frequency with an accuracy of order $T^{-4}$~\cite{lask99} (i.e., 
$\omega_1'-\omega_1=O(T^{-4})$) instead of an ordinary Fourier transform which 
gives an accuracy of order $T^{-1}$. The approximation of the next frequency 
$\omega_2$ is obtained by iterating the above procedure using $f_1(t)=f(t)-a'_1 
e^{i\omega'_1 t}$. Since the basis $\{ e^{i \omega'_k t} \}$ is not orthogonal 
for the scalar product $\langle \cdot  , \cdot \rangle$, a Gramm-Schmidt 
orthogonalization is necessary to obtain the next frequencies $\omega_k'$ for 
$k\geq 3$. \\  
     
For a chaotic trajectory, a ridge is a curve or a segment of curve 
$\xi_{loc}(u)$ in the time-frequency plane which is at each time $u$ a local 
maximum of the normalized scalogram, i.e.,
\begin{eqnarray*}
&& \left. \frac{\partial}{\partial \xi} P_W f(u,\xi) \right| _{\xi=\xi_{loc}(u)} 
=0, \\
&& \left. \frac{\partial^2}{\partial \xi^2} P_W f(u,\xi) \right| 
_{\xi=\xi_{loc}(u)} <0. 
\end{eqnarray*}      
Each ridge has a weight which is the value of the normalized scalogram on this 
ridge (it varies continuously in time). We will refer this value as the 
amplitude of the ridge. We call main ridge or main frequency, the ridge 
$\xi_m(u)$ (or a set of ridges) where the normalized scalogram is maximum:
$$
P_W f(u,\xi_m(u))=\max_\xi P_W f(u,\xi).
$$
In what follows, we will detect ridges by analyzing one coordinate of the 
system. We will show that the set of ridges of one coordinate contains all the 
frequency content of the trajectory, that is to say, the ridge plot will not be 
dependent on the choice of the coordinate used for the computation of the 
time-frequency transform. The respective amplitudes of the ridges depend on the 
specific coordinate chosen for the computation of the normalized scalogram. For 
instance, the main ridge depends in general on the 
chosen coordinate whereas the ridge plot does not.     \\
{\em Detection of the ridges.} For each time $u$, we are 
searching for the {\em local} maxima of the spectrogram or the normalized 
scalogram. In general, there is an infinite number of these ridges. We only 
detect the ones with sufficient amplitude according to a threshold $\epsilon$, 
i.e., we determine the set of ridges $\xi_{loc}(u)$ such that $$ P_W 
f(u,\xi_{loc}) \geq \epsilon \max_{\xi} P_W f(u,\xi). $$ 

     \section{Models}
\label{sec3}   
The time-frequency analysis can be performed on trajectories of 
maps or of flows.
In this  section, we illustrate the time-frequency analysis on the two 
models: the 
well-known standard map and the hydrogen atom in crossed magnetic and 
elliptically polarized microwave field.\\ The standard 
map is a two-dimensional area-preserving map   $(x,y)\mapsto (x',y')$~: 
\begin{eqnarray} && x'=x-a \sin y \\ && y'=x'+y \mod 2\pi.\end{eqnarray} This 
example provides a benchmark for the time-frequency analysis since the phase 
space structures are well-known in this example.\\ The hydrogen atom driven by a 
elliptically polarized microwave field (polarized in the orbital plane) and a 
constant magnetic field perpendicular to the orbital plane is described by the 
following Hamiltonian~\cite{Bfrie91}: \begin{equation} \label{eqn:hamcar} 
H(x,y,p_x,p_y,t)=\frac{p_x^2+p_y^2}{2}-\frac{1}{\sqrt{x^2+y^2}}+V_F(x,y,\omega 
t)+ V_B(x,y,p_x,p_y), \end{equation} where  $V_F$ is the contribution of the 
elliptically polarized microwave field and $V_B$ is the one of the magnetic 
field~: \begin{eqnarray*}                                                                 
&& V_F(x,y,\omega t)=F(x\cos \omega t + \alpha y \sin \omega t),\\ && 
V_B(x,y,p_x,p_y)=\frac{B}{2} (x p_y -y p_x)+\frac{B^2}{8}(x^2+y^2), 
\end{eqnarray*} where $\omega$ is the frequency of the microwave field and 
$\alpha$ is the ellipticity degree. For $\alpha=0$, the microwave field is 
linearly polarized (along the $x$ direction), and for $\alpha=1$, it is 
circularly polarized. In the circularly polarized case, the problem has two 
effective degrees of freedom. For $\alpha \in ]0,1[$, the field is elliptically 
polarized and is a three degree of freedom problem.  \\ The magnetic field 
$B$ perpendicular to the orbital plane of the electron limits the diffusion of 
trajectories~\cite{lee97,chan02b}. With it in place, trajectories will stay for 
a longer time close to the nucleus before ionizing, allowing longer trajectories 
to analyze by time-frequency analysis.

{\em Rescaling of the Hamiltonian:} By rescaling positions, momenta and 
time, we can assume that $\omega=1$. The rescaling of time is obtained by 
considering the Hamiltonian $\omega^{-1} H(x,y,p_x,p_y,t/\omega)$ where 
$H(x,y,p_x,p_y,t)$ is Hamiltonian~(\ref{eqn:hamcar}). Then we rescale positions 
and momenta~: $x'=\omega^{2/3} x$, $p'_x=\omega^{-2/3}p_x$, $y'=\omega^{2/3} y$ 
and $p'_y=\omega^{-2/3} p_y$, which is a canonical transformation generated by 
$S(x',y',p_x,p_y)=-\omega^{-2/3}(x'p_x+y'p_y)$. Then we rescale the momenta by 
changing $H$ into $\omega^{1/3} H(\omega^{-1/3} p_x, \omega^{-1/3} 
p_y,x,y)$. We notice that this transformation does not change the equations 
of motion. The resulting Hamiltonian is the same as 
Hamiltonian~(\ref{eqn:hamcar}) with $\omega=1$ and a rescaled value for the 
amplitude of the microwave field and the magnetic field: The amplitude of the 
microwave field is rescaled as $F'=F \omega^{-4/3}$ and the magnetic field as 
$B'=B/\omega$. In what follows, we assume that $\omega=1$.  

\subsection{Semi-parabolic coordinates}

The singularity of the potential in Hamiltonian~(\ref{eqn:hamcar}) is removed by
changing the coordinates to the semi-parabolic coordinates and a nonlinear
change of time. We perform the canonical transformation generated by
$$
S(u,v,p_x,p_y)=-\frac{u^2-v^2}{2}p_x -u v p_y.
$$                                                                      
The expression of the old coordinates as functions of the new coordinates is   
\begin{eqnarray*}
&& x=\frac{u^2-v^2}{2},\\
&& y= u v,\\
&& p_x=\frac{u p_u -v p_v}{u^2+v^2},\\
&& p_y =\frac{vp_u+up_v}{u^2+v^2}.
\end{eqnarray*}
The Hamiltonian becomes
\begin{eqnarray}
H=\frac{p_u^2+p_v^2}{2(u^2+v^2)} -\frac{2}{u^2+v^2} &+&F\left( 
\frac{u^2-v^2}{2}\cos t + \alpha  u v \sin t \right)\nonumber \\
 &+& \frac{B}{4}(u 
p_v -vp_u)+\frac{B^2}{32}(u^2+v^2)^2.   \label{eqn:hpart}
\end{eqnarray}         
We consider $t$ as a new variable $w$ with conjugate momentum $p_w$. The new 
Hamiltonian becomes time-independent:
\begin{eqnarray*}
H'=\frac{p_u^2+p_v^2}{2(u^2+v^2)} +p_w -\frac{2}{u^2+v^2} &+&F\left( 
\frac{u^2-v^2}{2}\cos w +\alpha u v \sin w \right)\\
 &+&\frac{B}{4}(u 
p_v -vp_u)+\frac{B^2}{32}(u^2+v^2)^2.
\end{eqnarray*}
We denote $E$ the energy of the system described by $H'$ (which is conserved 
since $H'$ does not depend on time).      Then we change time by multiplying the 
Hamiltonian by $u^2+v^2$. The new time $\tau$ satisfies $dt/d\tau =u^2+v^2$ 
(see Ref.~\cite{dela91}). We consider the effective Hamiltonian whose equations 
of motions are the same as the ones of Hamiltonian~(\ref{eqn:hpart}): 
\begin{eqnarray*} H_{eff}&=&\frac{p_u^2+p_v^2}{2}+(p_w-E) (u^2+v^2)-2\\ &&+F 
(u^2+v^2)\left ( \frac{u^2-v^2}{2}\cos  w +  \alpha u v \sin  w
\right )\\ && +\frac{B}{4}(u^2+v^2)(up_v-vp_u) +\frac{B^2}{32}(u^2+v^2)^3, 
\end{eqnarray*} 
where $H_{eff}=0$. We shift the momentum $p_w$ by a factor $-E$ and the 
effective Hamiltonian becomes \begin{eqnarray} \nonumber 
H_{eff}&=&\frac{p_u^2+p_v^2}{2}+ p_w (u^2+v^2)-2\\&& +F (u^2+v^2)\left ( 
\frac{u^2-v^2}{2}\cos w +  \alpha u v \sin w \right ) \nonumber \\ &&   
\label{eqn:hampar}    +\frac{B}{4}(u^2+v^2)(up_v-vp_u) 
+\frac{B^2}{32}(u^2+v^2)^3, \end{eqnarray}  We perform the following rotation 
of coordinates: \begin{eqnarray*} 
&&\tilde{u}= u\cos \frac{w}{2} +v\sin 
\frac{w}{2} ,\\ 
&&\tilde{v} = -u\sin \frac{w}{2} +v\cos \frac{w}{2}, \\
&&\tilde{w}=w,
\end{eqnarray*}                                                   
which is a canonical transformation generated by 
\begin{eqnarray*}
S(\tilde{u},\tilde{v},\tilde{w},p_u,p_v,p_w)=&&- p_u[\tilde{u}\cos
(\tilde{w}/2)-\tilde{v}\sin(\tilde{w}/2)] \\
&& -p_v[\tilde{u}\sin(\tilde{w}/2)+\tilde{v}\cos(\tilde{w}/2)]-\tilde{w} p_w.
\end{eqnarray*} The 
Hamiltonian~(\ref{eqn:hampar}) becomes \begin{eqnarray} 
\tilde{H}_{eff}=&&\frac{\tilde{p_u}^2+\tilde{p_v}^2}{2}+ \tilde{p_w} 
(\tilde{u}^2+\tilde{v}^2)-2 \nonumber \\
&& 
-\frac{1}{2}\left(1-\frac{B}{2}\right) 
(\tilde{u}^2+\tilde{v}^2)(\tilde{u}\tilde{ p _ v } - \tilde{v} \tilde{p_u} )+ 
\frac{F}{2} (\tilde{u}^4-\tilde{v}^4) \nonumber\\ && +(\alpha -1) F 
(\tilde{u}^2+\tilde{v}^2)\left[ \frac{\tilde{u}^2-\tilde{v}^2}{2}\sin \tilde{w} 
+\tilde{u}\tilde{v} \cos \tilde{w}\right] \sin \tilde{w} \nonumber  \\ &&  
+\frac{B^2}{32}(\tilde{u}^2+\tilde{v}^2)^3,  \label{hamrot} \end{eqnarray} where 
$\tilde{H}_{eff}$ is still equal to zero. In the circularly polarized case 
($\alpha=1$), the above Hamiltonian is independent of $\tilde{w}$, thus 
$\tilde{p_w}$ is constant, i.e., $p_w+(up_v-vp_u)/2=const$. Thus  the circularly 
polarized problem has two degrees of freedom. Poincar\'e surfaces of 
section will be drawn: the surfaces are the ones with $\tilde{P_\rho}=0$ (in the 
rotating frame), i.e., $\tilde{u}\tilde{p_u}+\tilde{v}\tilde{p_v}=0$.

\subsection{Choice of initial conditions}
We choose the initial conditions for the circularly polarized problem. Given the 
values of $F$ and $B$, the energy of the system is chosen according to 
Ref.~\cite{lee97} in the maximum configuration. At this energy, there is a 
stationary point of the flow and interesting dynamics happens in this case (see 
Ref.~\cite{lee97} for more details). We notice that the stationary points 
belong to the surface $\tilde{P}_{\rho}=0$. The maximum configuration is 
obtained for $\tilde{v}=0$, i.e.\ for \begin{equation} \label{eqn:pw} 
\tilde{p_w}=\frac{3}{8}(1-B)u_0^4-Fu_0^2, 
\end{equation}
where $u_0$ is the positive real solution of
$$
\frac{1}{2}(1-B) u_0^6 -F u_0^4-4=0.
$$
We notice that the equation that determines $u_0^2$ has always one positive real 
root and two complex conjugated roots for $F>0$ and $B<1$. \\
Once we have fixed $F$, $B$ and the energy of the system $p_w$, we compute 
Poincar\'e sections by the following procedure. These surfaces will be in the 
$(x,y)$-plane. We consider one point $(x,y)$ on this plane. We compute the 
associated $\tilde{u}$ and $\tilde{v}$:
\begin{eqnarray*}
&& \tilde{u}=\left(\sqrt{x^2+y^2}+x\right)^{1/2},\\
&& \tilde{v}=(\mbox{sgn } y) \left(\sqrt{x^2+y^2}-x\right)^{1/2}.
\end{eqnarray*}
The following conditions $\tilde{u}\tilde{p_u}+\tilde{v}\tilde{p_v}=0$ (which is 
$\tilde{P}_{\rho}=0$) and 
$\tilde{H}_{eff}(\tilde{u},\tilde{v},\tilde{p_u},\tilde{p_v})=0$ determine 
$\tilde{p_u}$ and $\tilde{p_v}$: \begin{eqnarray*}
&& \tilde{p_u}=-\frac{\tilde{v}}{2}\left[ \left(1-\frac{B}{2}\right) 
(\tilde{u}^2+\tilde{v}^2) +\varepsilon  \delta(\tilde{u},\tilde{v})\right] , \\ 
&& \tilde{p_v}=\frac{\tilde{u}}{2}\left[ 
\left(1-\frac{B}{2}\right) (\tilde{u}^2+\tilde{v}^2) 
+\varepsilon \delta(\tilde{u},\tilde{v}) \right], 
\end{eqnarray*} 
where $$ \delta(\tilde{u},\tilde{v})=\left[ 
(1-B)(\tilde{u}^2+\tilde{v}^2)^2-8\tilde{p_w}-4F(\tilde{u}^2-\tilde{v}^2)+16( 
\tilde{u}^2+\tilde{v}^2)^{-1}\right]^{1/2},   $$
and $\varepsilon=\pm 1 $ depending on whether we specify the trajectory 
crossing the Poincar\'e section $\tilde{P}_\rho=0$ in one direction 
($\dot{\tilde{P}}_\rho>0$) or the other ($\dot{\tilde{P}}_\rho<0$).    We use 
Poincar\'e sections in order to exhibit phase space structures and show that the 
time-frequency analysis of chaotic trajectories is able to reveal these 
structures.

\section{Illustration of the time-frequency method}
\label{sec4}
In what follows, we consider the trajectories of the standard map (discrete 
time) and the trajectories of the hydrogen atom in crossed fields (continuous 
time).  We will show that the phase space structures encountered by a trajectory
are revealed by the time-frequency analysis of a single coordinate of this 
trajectory, i.e., by the computation of the ridges of the time-frequency 
landscape, the instantaneous frequencies.  \\ We compute the time-frequency 
content of real time series (e.e., one coordinate of the system) obtained by 
iterating the map or integrating the flow. We only compute the time-frequency 
content of the trajectory in the positive frequency region.

\subsection{Trajectories of the Standard Map}
\subsubsection{Quasiperiodic trajectory}
\begin{figure}
\centerline{\includegraphics*[scale=0.5]{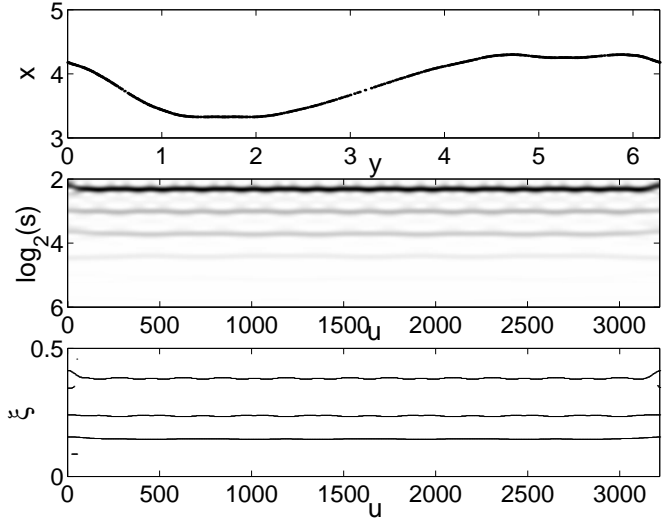}}       
\caption{\label{fig1} Phase portrait (first panel), normalized scalogram (second 
panel) and ridge plot (third panel) of a quasiperiodic trajectory of the 
standard map obtained for $K=0.971$ and initial conditions $x_0=4.176562$ and
$y_0=0$.}
\end{figure}

We consider a trajectory of the standard map for $K=0.9710$
starting at $x_0=4.176562$ and $y_0=0$ over a time span of
512 iterations (which corresponds to $T=3217$ for the corresponding flow). 
This trajectory, plotted in the first panel of Fig.~\ref{fig1}, corresponds to 
the invariant torus (golden mean torus) whose break-up is studied in 
Ref.~\cite{lask92} using frequency map analysis.  The second panel of 
Fig.~\ref{fig1} depicts the normalized scalogram in the time-scale plane, 
i.e.\ the amplitude of the wavelet coefficients, obtained from the $x$ 
coordinate. From this time-scale (or time-frequency) landscape, we extract 
the local maxima which define the ridges or the instantaneous frequencies of 
the trajectory. The third panel of Fig.~\ref{fig1} shows the different ridges 
of the normalized scalogram (the ridge detection threshold is $\epsilon=0.1$). 
These ridges are approximately constant in time as one expects. The upper 
curve represents the main frequency
$\omega_0=(3-\sqrt{5})/2$.  The other two main frequency 
curves in the ridge plot are located at $\omega_1=1-2\omega_0$ and 
$\omega_2=3\omega_0-1$.  For a system with two degrees of freedom, there are at 
most two independent frequencies: in this case, they are equal to $\omega_0$ and 
1 which is the natural frequency for constructing the Poincar\'e map from the 
flow (kicked rotor). Each ridge (a priori an infinite number) is a combination 
of these two frequencies.

We can also see the interferences (oscillations around the
expected frequency) between these frequencies. In order to reduce
these oscillations, we can increase the parameter $\eta\sigma$. We notice that 
the amplitudes or frequencies of these oscillations do not depend on the length 
of the time interval in which the trajectory is computed (see Appendix A).   

\subsubsection{Weakly chaotic motion}

\begin{figure}
\centerline{\includegraphics*[scale=0.5]{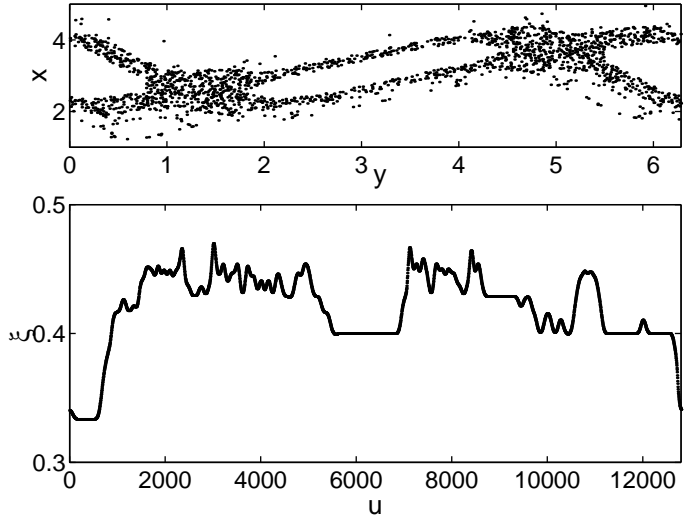}}
 \caption{\label{fig2} Phase portrait (first panel) and ridge plot (second panel)
  of  a weakly chaotic trajectory of the
standard map obtained for $K=1.2$ and initial conditions $x_0=2.5$ and
$y_0=\pi$.}
\end{figure}

We consider a trajectory obtained for the standard map with
$K=1.2$ and initial conditions $x_0=2.5$ and $y_0=\pi$. We analyze
this weakly chaotic trajectory over the time interval of total
length $T=12868$. The upper panel of Fig.~\ref{fig2} depicts this trajectory. 
The lower panel depicts the main ridge (the ridge detection threshold is 
$\epsilon=0.5$) and shows clearly different trappings as time evolves: trappings 
in the resonance 1:3 for $150\leq t \leq 500$, in 2:5 for $5600 \leq t\leq 
6800$, in 4:7 for $8700 \leq t \leq 9300$, and in 3:5 for $11200\leq t \leq 
12600$. These trappings result from the trajectory passing nearby islands 
surrounding elliptic periodic orbits (the other ridges contain 
the secondary frequencies of the nearby quasiperiodic motion). The transitions
between these resonances occur quite smoothly, passing by other
resonance trappings. They occur when the chaotic trajectory passes nearby
hyperbolic periodic orbits where great variations of the frequencies are
expected. The trajectory is weakly chaotic since the main
information can be obtained from a single instantaneous
frequency curve. In this regime, looking at the frequency where
the spectrogram or scalogram is maximum (main instantaneous frequency) is 
meaningful.                                                       

\subsubsection{Resonance transitions}
\begin{figure}
\centerline{\includegraphics*[scale=0.5]{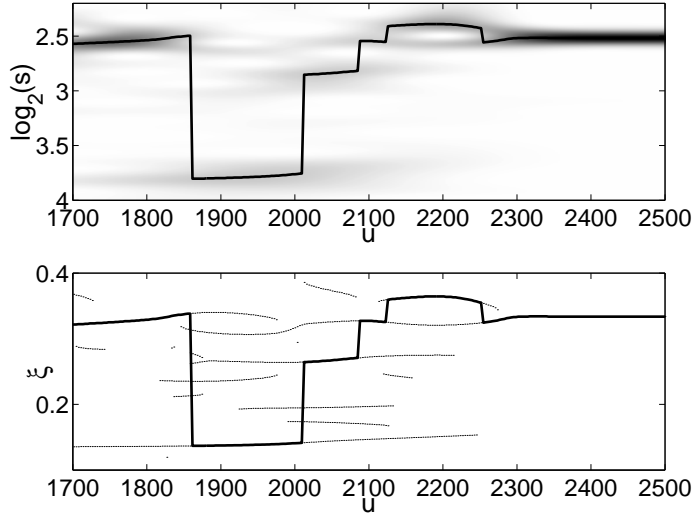}}
 \caption{\label{fig3} Ridge plot of the $x$ component of the trajectory of the
standard map with $K=1.5$ and initial conditions $x_0=1.35$ and $y_0=0$ on the 
time interval $[1700,2500]$. The upper panel is the time-scale plane and the 
lower panel is the ridge plot. The bold line denotes the main frequency curve.} 
\end{figure}

The time-frequency resolution of wavelets allows us to analyze the
mechanism of resonance transition. Figure~\ref{fig3} shows the normalized 
scalogram in the time-scale plane and the ridge plot (with ridge detection 
threshold $\epsilon=0.1$) obtained for the $x$ coordinate of a trajectory of the 
standard map with $K=1.5$ and initial conditions $x_0=1.35$ and $y_0=0$ on a 
time interval $[1700,2500]$. In this figure, the bold curve represents 
the main ridge or the main frequency (i.e.\ where the normalized 
scalogram is maximum, see Sec.\ \ref{sec24}).\\

In this time interval, several transitions occur. For instance around $t=1860$, 
the main frequency jumps discontinuously from the ridge located around 
$\omega_1\approx 0.33$ to the ridge located around $\omega_2\approx 0.136$. One 
could wrongly conclude that a sharp resonance transition occurred in the system. 
By looking at secondary ridges, i.e.\ the ridges with smaller amplitudes than 
the main frequency, one can see that both ridges existed before and after 
$t=1860$; Nothing discontinuous happens in the system. From the ridge plot, the 
mechanism of resonance transition appears to be the following one: A second 
ridge at $\xi=\omega_2$ increases in amplitude from $t=1700$ to $t=1860$ while 
the first ridge at $\xi=\omega_1$ decreases in amplitude. On some time interval 
(from $t=1800$ to $t=2000$), several ridges of the same order in amplitude 
appear. At $t=1860$, the second ridge becomes the main frequency.   This 
situation is typical from other ridge plots of trajectories showing transitions 
from ridges and hence resonance transitions (see Fig.~\ref{fig12}). Thus we 
have a new characterization of resonance transition that applies in more general 
situations. Resonance transitions are a manifestation of the fulfillment of the 
Chirikov resonance overlap criterion~\cite{chir79}. 

\subsubsection{Strongly chaotic motion}         
\begin{figure}
\centerline{\includegraphics*[scale=0.5]{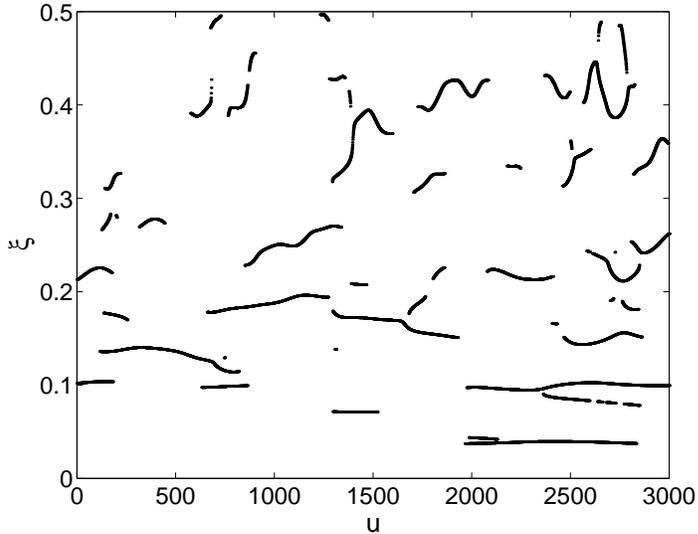}}
 \caption{\label{fig4} Ridge plot of a strongly chaotic trajectory of the
standard map with $K=5$ and initial conditions $x_0=0.95$ and $y_0=0$.}
\end{figure}

When $K$ increases, the trajectories are increasingly chaotic (filling a bigger 
part of phase space). More and more short ridges appear, and in those
cases, looking at the main frequency leads to erroneous results
as shown above since several ridges have the same amplitude. 
The time-frequency information cannot be reduced to (or deduced from) a single
instantaneous frequency. For instance, the main frequency curve could follow 
a short flat ridge and one could wrongly deduce a trapping region. 
We consider a strongly chaotic trajectory of the standard map
obtained with $K=5$ and initial conditions $x_0=0.95$ and $y_0=0$.
Figure~\ref{fig4} shows the ridge plot obtained from the
normalized scalogram (with a ridge detection threshold $\epsilon=0.5$). It shows 
that no trappings occur in this time interval. The presence of a lot (but a 
finite number) of ridges reflects the broad-band spectrum of a chaotic 
trajectory. These pictures lead to a characterization of weak and strong chaos
by looking at the number of important ridges in the system: Weak
chaos is characterized by one main connected instantaneous
frequency curve whereas strong chaos is characterized by multiple
short ridges (the number of them increases as $K$ increases).

\subsection{Hydrogen atom in crossed fields}
\begin{figure}
\centerline{\includegraphics*[scale=0.5]{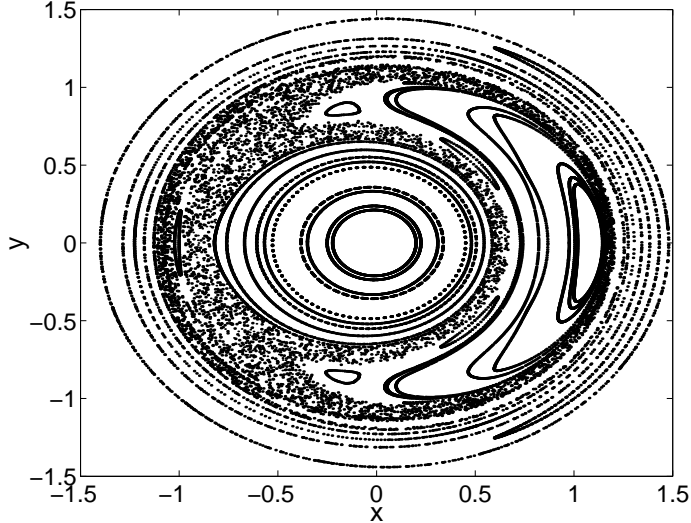}}
 \caption{\label{fig5} Poincar\'e section of the 
Hamiltonian~(\ref{hamrot}) in the maximum 
configuration for $F=0.015$ and $B=0.3$.} \end{figure}

We consider in this section the trajectories of the Hamiltonian flow of the 
hydrogen atom driven by crossed magnetic and elliptically polarized microwave 
field in the rotating frame represented by Hamiltonian~(\ref{hamrot}) in the 
maximum configuration.\\ Otherwise specified, we consider in this section the 
following values for the parameters: $F=0.015$ and $B=0.3$. Figure~\ref{fig5} 
depicts a Poincar\'e section of the Hamiltonian in the circularly polarized 
case. We analyze the trajectories by applying the wavelet transform on one real 
coordinate $x(t)$ or $y(t)$.       

\subsubsection{Periodic orbit}    
\begin{figure} 
\centerline{\includegraphics*[scale=0.5]{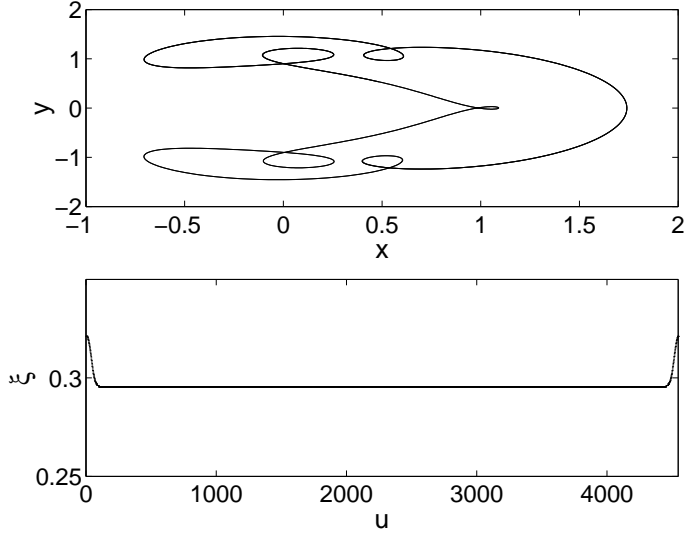}}  \caption{\label{fig6} Ridge 
plot of a periodic orbit of Hamiltonian~(\ref{hamrot}) for $F=0.015$ and $B=0.3$ 
in the maximum configuration obtained for initial conditions 
$(x,y)=(0.4835,0.5335)$ in the Poincar\'e section. The upper part of the figure 
represents a projection of the periodic orbit on the $x-y$ plane.} \end{figure}

We consider a closed trajectory in phase space obtained for initial conditions 
$(x,y)\approx (0.4835,0.5335)$ in the Poincar\'e section.  A projection of this periodic 
orbit on the plane $(x,y)$ is shown in the upper part of Fig.~\ref{fig6}. The 
ridge plot of this orbit is depicted in the lower panel of Fig.~\ref{fig6}. \\ 
The ridge plot displays only one flat ridge located at the value of its 
frequency $\omega_A\approx 0.2952$.  The harmonics are not visible on the ridge 
plot since these ridges are too small compared to the main ridge (ridge 
detection threshold $\epsilon=0.1$).  
The bumps at $u=0$ and $u=4500$ are a consequence of the finite time interval on 
which the periodic trajectory is computed. This effect is discussed in Appendix 
B.  \\ There are other elliptic periodic orbits in the system: For instance, the 
one obtained from the initial conditions in the Poincar\'e section 
$(x,y)=(-1.014,0)$ with frequency $\omega_B\approx0.3033$, and for 
$(x,y)=(-0.807,0.558)$ with frequency $\omega_C\approx 0.2260$.  

\subsubsection{Quasiperiodic trajectory}    
\begin{figure} 
\centerline{\includegraphics*[scale=0.5]{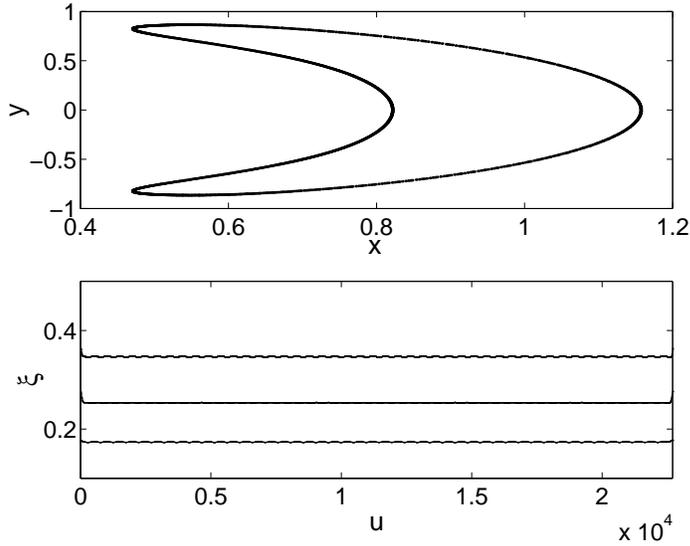}}  \caption{\label{fig7} Ridge 
plot of a quasiperiodic orbit of Hamiltonian~(\ref{hamrot}) for $F=0.015$ and 
$B=0.3$ in the maximum configuration obtained for initial 
conditions $(x,y)=(0.7,0.5)$ in the Poincar\'e section. The upper part of the 
figure represents the Poincar\'e section of the trajectory.} \end{figure}

The ridge plot of a quasiperiodic orbit obtained for initial conditions 
$(x,y)=(0.7,0.5)$ is depicted in Fig.~\ref{fig7}. It displays a set of three 
main ridges around $\omega_1\approx 0.17$, $\omega_2\approx 0.25$ and 
$\omega_3\approx 0.34$ (with a ridge detection threshold $\epsilon=0.1$). The 
oscillations are due to the interactions between ridges. This effect is 
explained in Appendix A.     
The frequency map 
analysis gives the following values for the first frequencies:  
$\omega_1=0.1735$, $\omega_2=0.2531$ and 
$\omega_3=2\omega_1=0.3470$.   

\subsubsection{Chaotic orbits: resonance trappings and transitions}

\begin{figure} 
\centerline{\includegraphics*[scale=0.5]{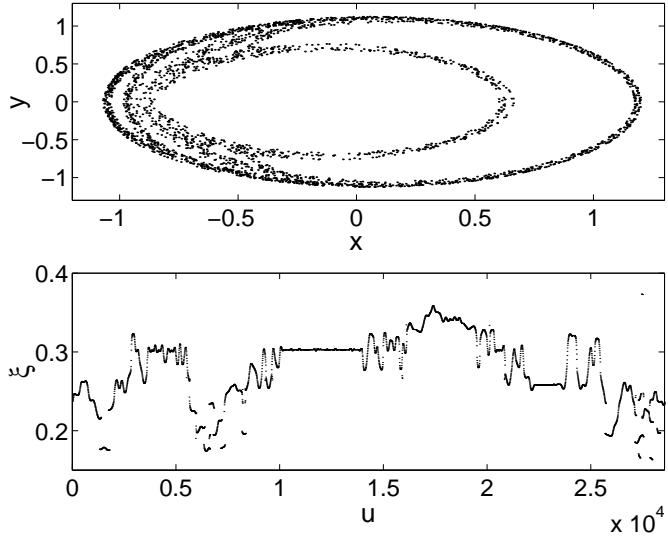}}  \caption{\label{fig8} 
Ridge plot of a chaotic orbit of Hamiltonian~(\ref{hamrot}) for $F=0.015$ and 
$B=0.3$ in the maximum configuration obtained for initial conditions: 
$(x,y)=(-0.5,0.7)$. The upper part of the figure represents the Poincar\'e 
section of the trajectory.} \end{figure}

We consider a trajectory in a chaotic region obtained for initial conditions 
$(x,y)=(-0.5,0.7)$ in the Poincar\'e section. Figure~\ref{fig8} depicts a 
Poincar\'e section of this trajectory and the ridge plot of the coordinate 
$x(t)$ over the time interval $ [0,2.85~ 10^{4}]$ (with a ridge detection 
threshold $\epsilon=0.5$).            

\begin{figure} \centerline{\includegraphics*[scale=0.5]{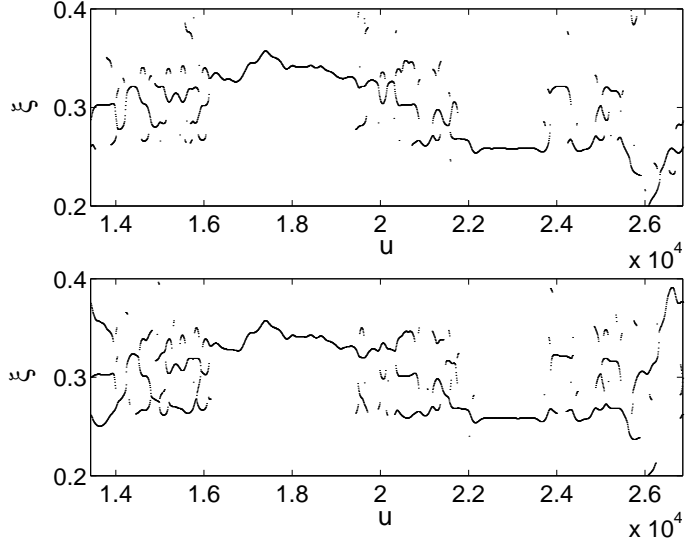}}  
\caption{\label{fig9} Ridge plot of the chaotic orbit of Fig.~\ref{fig8} on 
the time interval $[1.35~10^4, 2.65~10^4]$. The upper part of the ridge plot is 
obtained by analyzing the coordinate $x(t)$ and the lower part is the one for 
the coordinate $y(t)$.} \end{figure}

Figure~\ref{fig9} depicts an enlargement of the ridge plot of Fig.~\ref{fig8} 
on the time interval $[1.35~ 10^4, 2.65 ~10^4]$ of two coordinates
$x$ and $y$ with a smaller ridge detection threshold $\epsilon=0.1$. Similar 
ridge plots are obtained with the other coordinates $p_x$ and $p_y$.  

\begin{figure} 
\centerline{\includegraphics*[scale=0.5]{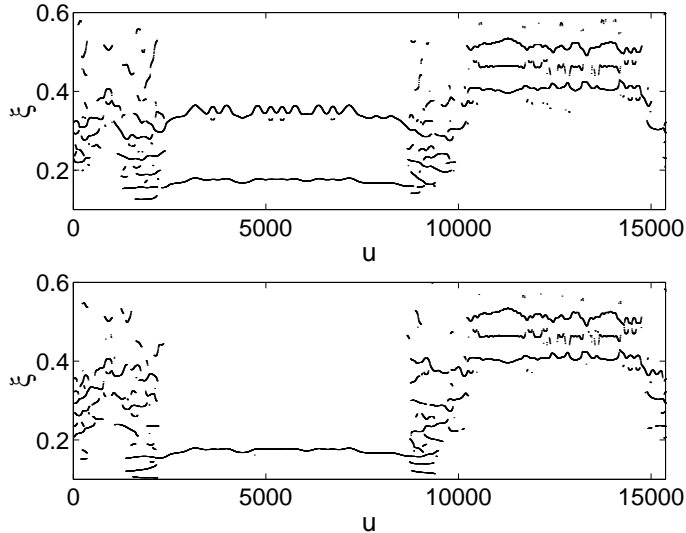}}  \caption{\label{fig10} 
Ridge plot of the chaotic orbit of Hamiltonian~(\ref{hamrot}) in the 
maximum configuration with the parameters $F=0.02$ and $B=0.3$  obtained for 
initial conditions $(x,y)=(-0.7,-0.5)$ in the Poincar\'e section. The upper part 
of the ridge plot is obtained by analyzing the coordinate $x(t)$ and the lower 
part is obtained from the coordinate $y(t)$.} \end{figure}

\begin{figure} 
\centerline{\includegraphics*[scale=0.5]{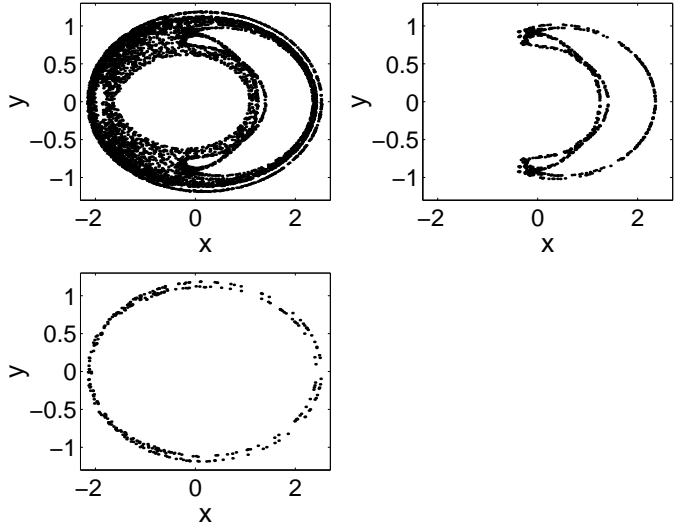}}  
\caption{\label{fig11} Poincar\'e section of the trajectory of 
Fig.~\ref{fig10}. The upper left panel is the trajectory from $t=0$ to $t=6.1 
~10^4$, the upper right panel is the trajectory from $t=3~10^3$ to $t=8~10^3$, 
and the lower panel is the trajectory from $t=10.4 ~10^3$ to $t=14.5~10^3$.} 
\end{figure}

Three different type of regimes are observed on these 
figures: trapping regions 
where the frequency is constant, weakly chaotic regions 
where the frequency is changing smoothly (for instance, from $t=1.6~ 10^4$ to 
$t=1.95 ~10^4$), and strongly chaotic regions characterized by many 
short ridges (for instance between $t=5.5~ 10^3$ and $t=8.5 ~10^3$).\\ 
The two main trapping regions in these figures are obtained from $t=10^4$ to 
$t=1.38~ 10^4$ where the trajectory is trapped around the elliptic island 
surrounding the periodic orbit with frequency $\omega_B$, and from $t=2.23~ 
10^4$ to $t=2.35~ 10^4$ where the trajectory is trapped around the elliptic 
island with frequency $\omega_C$.\\ The part of phase space visited by the 
trajectory is compact (this is due to the magnetic field) so we do not expect 
great variations of the main frequency.\\ In Fig.~\ref{fig10}, we depict a 
resonance transition that occurs in the system for $F=0.02$ and $B=0.3$ (still 
in the maximum configuration) for initial conditions $(x,y)=(-0.7,-0.5)$ in the 
Poincar\'e section. This figure displays the ridges of the coordinates $x$ and 
$y$ with a ridge detection threshold $\epsilon=0.1$.  There are two 
trapping regions: from $t=3000$ to $t=8000$ and from $t=10400$ to $t=14500$. The 
first trapping region is characterized by two main frequencies $\omega_1=0.1750$ 
and $\omega_2=2\omega_1$, and the second trapping region has three main 
frequencies $\omega'_1=0.4106$, $\omega_2'=0.4630$ and 
$\omega'_3=2\omega_2-\omega_1$.  Both regions are separated by a strongly 
chaotic region characterized by  many short ridges (resonance transition 
region). Figure~\ref{fig11} displays Poincar\'e section of the trajectory from 
$t=0$ to $t=61400$ (upper left panel), the section of this trajectory which is 
trapped between $t=3000$ and $t=8000$ (upper right panel), and the section of 
the trajectory trapped from $t=10400$ to $t=14500$ (lower panel). In the trapped 
regions, the trajectory nearly fills a one-dimensional part of the Poincar\'e 
surface.

\begin{figure} 
\centerline{\includegraphics*[scale=0.5]{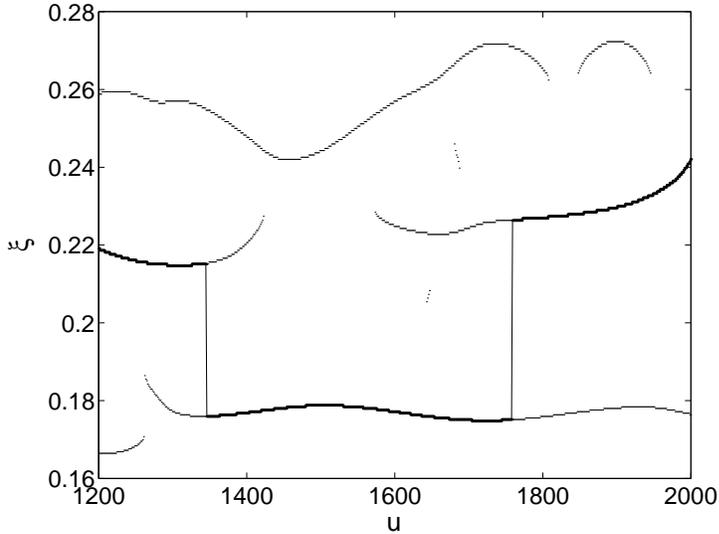}}  
\caption{\label{fig12} Ridge plot of the same trajectory as for 
Fig.~\ref{fig8} on the time interval $[1200, 2000]$. The bold curve represents 
the main frequency curve.} \end{figure}

From Figs.~\ref{fig9} and \ref{fig10}, we notice that the ridge plots are the 
same for the $x$ coordinate of the $y$ coordinate. The only difference is that a 
given ridge has different amplitudes according to whether it has been computed 
using $x(t)$ or $y(t)$. One exception is 
Fig.~\ref{fig10} from $t=3000$ to $t=8000$: only the ridges of sufficient 
amplitude are detected (with a ridge detection threshold $\epsilon=0.1$) 
and the ridge around $\xi\approx 0.35$ which is the main ridge for the $x$ 
coordinate has a low amplitude for the $y$ coordinate. This ridge appears if we 
include low amplitude ridges in the detection.\\ 
The noticeable differences 
between the ridge plots for $x$ and the ones for $y$ are in the strong chaotic 
region where there are no elliptic islands and thus no trappings. In this regime 
which is the same as the one in Fig.~\ref{fig4} for the standard map, the 
significance of a frequency or a set of frequencies as an indicator is not 
clear. It means that the basis is not adapted to the strong chaotic regime. 
However, computing the ridge curves is useful since it provides a way to locate 
these strongly chaotic regions. It also provides a quantitative distinction 
between weak and strong chaos by looking at the number of short ridges.\\  
We highlight that computing a single instantaneous frequency from the 
maximum wavelet coefficients can provide erroneous information. 
Figure~\ref{fig12} depicts the ridges in a strongly chaotic zone of phase 
space (the same trajectory as for Fig.~\ref{fig8} on the time interval $[1200, 
2000]$). The bold curve is the main frequency ridge. By looking at this 
main frequency, we cannot say if the trajectory is in a strong chaotic 
region or if it sharply jumps between different trapping regions. The 
computation of secondary ridges shows that it is indeed a strongly chaotic zone 
without trappings.  

\subsubsection{Time-frequency analysis of higher dimensional systems}

\begin{figure} 
\centerline{\includegraphics*[scale=0.5]{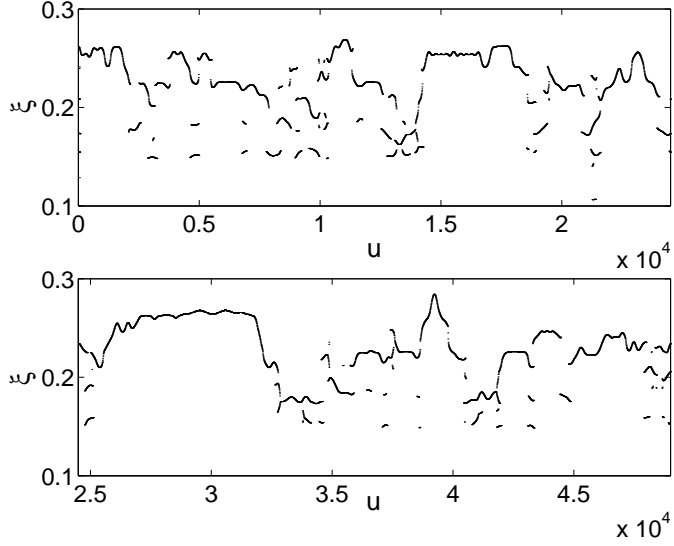}}  
\caption{\label{fig13} Ridge plot of  the $x$ coordinate of the chaotic orbit of 
Hamiltonian~(\ref{hamrot}) with $\alpha=0.9$ with 
the parameters $F=0.015$ and $B=0.3$  obtained for initial conditions 
$(x,y)=(-0.5,0.7)$ and $p_w$ solution of Eq.~(\ref{eqn:pw}); $p_x$ and $p_y$ are 
solution of $\tilde{H}_{eff}=0$ and $xp_x+yp_y=0$. The upper part of the ridge 
plot is for the time interval $[0, 2.4~10^4]$ and the lower part is for 
$[2.4~10^4, 4.9~10^4]$. The ridge detection threshold is $\epsilon=0.5$.} 
\end{figure}

\begin{figure} 
\centerline{\includegraphics*[scale=0.5]{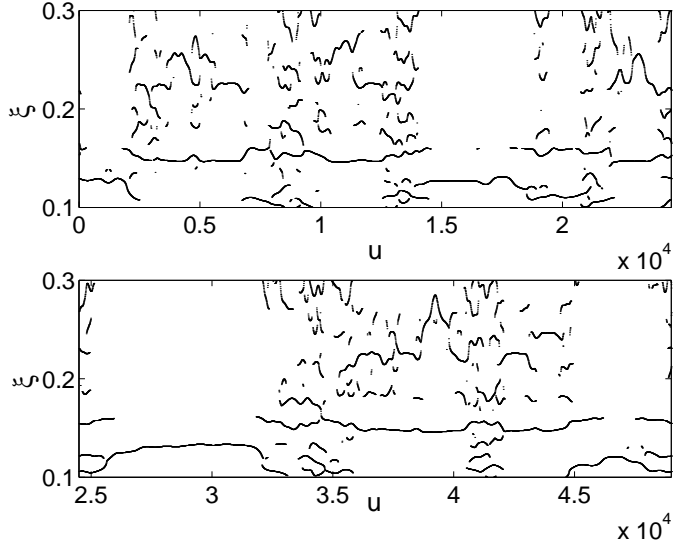}}  
\caption{\label{fig14} Ridge plot of  the $y$ coordinate of the same trajectory 
as Fig.~\ref{fig13} with a ridge detection threshold $\epsilon=0.1$.} 
\end{figure}

The time-frequency analysis can be carried out for higher dimensional systems. 
We consider the elliptically polarized case with $\alpha=0.9$ represented by 
Hamiltonian~(\ref{hamrot}) in the rotating frame. In this case, $\tilde{p_w}$ is 
no longer a conserved quantity so Hamiltonian~(\ref{hamrot}) has three degrees 
of freedom. Figure~\ref{fig13} depicts the ridge plot of the $x$ coordinate with 
a ridge detection threshold $\epsilon=0.5$, and Figure~\ref{fig14} depicts the 
ridge plot of the $y$ coordinate with a ridge detection threshold 
$\epsilon=0.1$. 

\begin{figure} 
\centerline{\includegraphics*[scale=0.5]{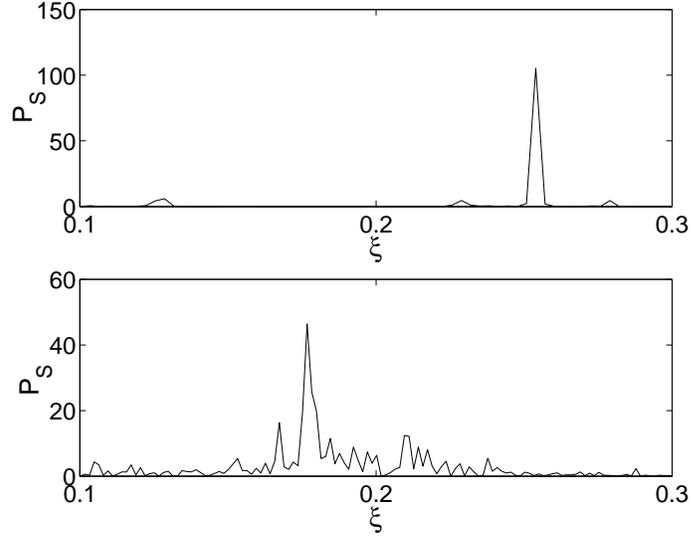}}  \caption{\label{fig15} 
Power spectrum of two segments of the trajectory of Fig~\ref{fig13} ($x$ 
coordinate): from $t=1.45~10^4$ to $t=1.65~10^4$ (upper panel), and  from 
$t=3.2~10^4$ to $t=3.6~10^4$ (lower panel).} \end{figure}

 We notice that from this figure the trajectory looks more chaotic 
than in the circularly polarized case which is in agreement with 
Refs.~\cite{oks99,bell97} (see the differences between Fig.~\ref{fig8} and 
Fig.~\ref{fig13}). From Figs.~\ref{fig13} and \ref{fig14}, we detect resonance 
trappings and weakly chaotic regions. For instance, from $t=1.45~10^4$ to 
$t=1.65 ~10^4$ the trajectory is trapped in a region where the main frequencies 
are $\omega_1=0.127$ (which is the main frequency for $y$) and 
$\omega_2=2\omega_1$ (which is the main frequency for $x$). The same trapping is 
observed on a longer time interval from $t=2.65~10^4$ to $t=3.15~10^4$. This is 
confirmed by a Fourier analysis of the segments of the trajectory: 
Figure~\ref{fig15} shows the power spectrum of two segments of the trajectory: 
from $t=1.45~10^4$ to $t=1.65~10^4$ which is expected to be close to 
quasiperiodic, and from $t=3.2~10^4$ to $t=3.6~10^4$ which is expected to be 
strongly chaotic with a broad-band spectrum.  

By increasing the ridge detection threshold (see Fig.~\ref{fig14}), we clearly 
see the trapping regions (with one main flat ridge) and the strongly chaotic 
regions. We notice that there are regions where there is one main flat ridge 
with additional short ridges (for instance from $t=10^4$ to $t=1.25~10^4$ or 
from $t=3.6~10^4$ to $t=4~10^4$). A Fourier analysis of these segments of 
trajectory shows a main frequency $\omega\approx 0.15$ 
and a broad-band component. This is a nearly trapped region near some elliptic 
island. From Fig.~\ref{fig14}, we notice that the trajectory is trapped in this 
island from $t=4.22~10^4$ to $t=4.3~10^4$, and the main frequencies are 
$\omega_1=0.15$ and $\omega_2=0.22$.

\section{Conclusion}
In summary, we showed how instantaneous frequencies based on a ridge extraction 
of a wavelet decomposition of a single trajectory coordinate 
reveal the phase space structures (resonance transitions, trappings, etc.) of 
chaotic systems. Our method also gives a quantitative characterization of weak 
and strong chaos through trappings and through the number of short ridges. We 
have showed that this time-frequency analysis can be carried out on trajectories 
of maps and flows. Note that nothing in our
analysis depends on the dimensionality of the system: Given a time
series (which can be e.g., one of the coordinates of the system
whatever the number of such coordinates may be, or indeed data sets), we
can compute the ridges of the time-frequency landscape, leading to the 
instantaneous frequencies. Moreover, the computational time is 
independent of the dimensionality of the system. The time-frequency method 
based on a single trajectory analysis is therefore very well suited for 
exploring phase space structures of various systems independently of the number 
of degrees of freedom.   

\section*{Acknowledgments}
CC acknowledges useful discussions with N. Garnier and J. Laskar.  This work was 
supported by the National Science Foundation and by ONR Grant No.~N00014-01-1-0769.

\appendix
\section{Time-frequency analysis of quasiperiodic trajectories}
Given the following quasiperiodic function
\begin{equation}
\label{eqn:qps}
f(t)=e^{i\omega_1 t} +\sum_{k>1} a_k e^{i\omega_k t},
\end{equation}
where $a_k\in \mathbb{C}$, we derive an approximation of the ridges of the 
time-frequency landscape obtained using the windowed Fourier transform and the 
continuous wavelet transform. 
\subsection{Windowed Fourier transform}
The windowed Fourier transform of  trajectory (\ref{eqn:qps}) is (up to an 
overall constant $\sigma\sqrt{2\pi}$): $$
Sf(u,\xi)=e^{i\omega_1 u} e^{-\sigma^2(\omega_1-\xi)^2/2} +\sum_{k>1} a_k 
e^{i\omega_k u} e^{-\sigma^2(\omega_k-\xi)^2/2}.
$$
We want to study the behavior of $Sf(u,\xi)$ around $\xi\approx \omega_1$. 
We expand $Sf(u,\xi)$ up to second order in the variable 
$x=\xi-\omega_1$~:
$$
Sf(u,\xi)=Sf^{(0)} +Sf^{(1)} x + Sf^{(2)} x^2 +O(x^3),
$$
where
\begin{eqnarray*}
&& Sf^{(0)}= e^{i\omega_1 u} +\sum_{k>1} a_k e^{i\omega_k u} 
e^{-\sigma^2(\omega_k -\omega_1)^2/2},\\
&& Sf^{(1)}=\sigma^2 \sum_{k>1} a_k (\omega_k-\omega_1) e^{i\omega_k u} 
e^{-\sigma^2(\omega_k-\omega_1)^2/2},\\
&& Sf^{(2)}=-\frac{1}{2}\sigma^2 \left\{ e^{i\omega_1 u}+\sum_{k>1} a_k 
[1-\sigma^2(\omega_k-\omega_1)^2 ] e^{i\omega_k u} 
e^{-\sigma^2(\omega_k-\omega_1)^2/2} \right\}.
\end{eqnarray*}
The ridge around $\omega_1$ is the local maximum of the absolute value 
of $Sf(u,\xi)$ for each $u$. The expression of the spectrogram is 
\begin{eqnarray*}
|Sf(u,\xi)|^2=\vert Sf^{(0)}\vert^2 &+&2\mbox{Re}\left( Sf^{(0)} 
Sf^{(1)*}\right ) x\\
 &+& \left [ 2\mbox{Re}\left( Sf^{(0)} 
Sf^{(2)*}\right ) +\vert Sf^{(1)}\vert ^2\right ] x^2 +O(x^3).
\end{eqnarray*}
The ridge is approximately located at 
$$
\xi(u)=\omega_1-\frac{  \mbox{Re}\left( Sf^{(0)} 
Sf^{(1)*}\right )}{2      \mbox{Re}\left( Sf^{(0)} 
Sf^{(2)*}\right )  +     \vert Sf^{(1)}\vert ^2}.
$$
When the terms $a_k e^{-\sigma^2(\omega_k-\omega_1)^2/2}$ are small (compared to 
1) which is the case when $a_k$ is small or $\omega_k$ is sufficiently far from 
$\omega_1$, the expression of the ridge becomes~: \begin{equation}
\xi(u)=\omega_1+\sum_{k>1} (\omega_k-\omega_1) A_k \cos [(\omega_k-\omega_1) u 
+\phi_k ]  e^{-\sigma^2(\omega_k-\omega_1)^2/2},
\end{equation}
where $a_k=A_k e^{i\phi_k}$.
The instantaneous frequency curve is again a quasiperiodic function with 
frequencies $\omega_k-\omega_1$. It oscillates around the main frequency 
$\omega_1$ with amplitude proportional to 
$e^{-\sigma^2(\omega_k-\omega_1)^2/2}$. These oscillations can be reduced by 
increasing the length of the window of the transform $\sigma$.

\subsection{Wavelet transform}
     The wavelet transform of  trajectory (\ref{eqn:qps}) is (up to an overall 
constant $\sigma \sqrt{2\pi}$): $$
Wf(u,s)=e^{i\omega_1 u} e^{-\sigma^2 s^2(\omega_1-\eta/s)^2/2} +\sum_{k>1} a_k 
e^{i\omega_k u} e^{-\sigma^2 s^2(\omega_k-\eta/s)^2/2}.
$$
We want to study the behavior of the normalized scalogram $P_W f (u,\xi)$ around 
$s\approx \eta/\omega_1$. We expand $s^{-1/2} Wf(u,s)$ up to second order 
in the variable $x=s\omega_1/\eta-1$~:
$$
s^{-1/2}Wf(u,s)=Wf^{(0)} +Wf^{(1)} x + Wf^{(2)} x^2 +O(x^3),
$$
where
\begin{eqnarray*}
&& Wf^{(0)}= e^{i\omega_1 u} +\sum_{k>1} a_k e^{i\omega_k u} 
e^{-\sigma^2\eta^2(\omega_k/\omega_1 -1)^2/2},\\
&& Wf^{(1)}=-\sigma^2\eta^2 \sum_{k>1} a_k 
\frac{\omega_k}{\omega_1}\left(\frac{\omega_k}{\omega_1}-1\right ) e^{i\omega_k 
u} e^{-\sigma^2\eta^2(\omega_k/\omega_1-1)^2/2},\\ && 
Wf^{(2)}=-\frac{1}{2}\sigma^2\eta^2 \left\{ e^{i\omega_1 u}+\sum_{k>1} a_k 
\frac{\omega_k^2}{\omega_1^2}\left[ 1-\sigma^2\eta^2\left( 
\frac{\omega_k}{\omega_1}-1\right)^2 \right] e^{i\omega_k u} 
e^{-\sigma^2\eta^2(\omega_k/\omega_1-1)^2/2} \right\}. \end{eqnarray*} The ridge 
around $\omega_1$ is given by the local maximum of 
$P_W f (u,\xi)$ for each $u$. The expression of the normalized scalogram 
\begin{eqnarray*}
\frac{1}{s}|Wf(u,s)|^2=\vert Wf^{(0)}\vert^2 &+&2\mbox{Re}\left( Wf^{(0)} 
Wf^{(1)*}\right ) x \\ &+& \left [ 2\mbox{Re}\left( Wf^{(0)} 
Wf^{(2)*}\right ) +\vert Wf^{(1)}\vert ^2\right ] x^2 +O(x^3).
\end{eqnarray*}
The ridge is approximately located at 
$$
\xi(u)=\omega_1-\frac{  \mbox{Re}\left( Wf^{(0)} 
Wf^{(1)*}\right )}{2      \mbox{Re}\left( Wf^{(0)} 
Wf^{(2)*}\right )  +     \vert Wf^{(1)}\vert ^2}.
$$
When the terms $a_k e^{-\sigma^2\eta^2(\omega_k/\omega_1-1)^2/2}$ are small 
(compared to 1) which is the case when $a_k$ is small or $\omega_k$ is 
sufficiently far from $\omega_1$, the expression of the ridge becomes~: 
\begin{equation} \xi(u)=\omega_1+\sum_{k>1} \omega_k(\omega_k/\omega_1-1) A_k 
\cos [(\omega_k-\omega_1) u +\phi_k ]  
e^{-\sigma^2\eta^2(\omega_k/\omega_1-1)^2/2}. \end{equation}

As for the windowed Fourier transform, the instantaneous frequency is a 
quasiperiodic function with frequencies $\omega_k-\omega_1$. It oscillates 
around the main frequency $\omega_1$. The amplitude of the oscillations is 
proportional to $e^{-\sigma^2\eta^2(\omega_k/\omega_1 -1)^2/2}$. They can be 
reduced by increasing the parameters of the wavelet $\sigma$ and $\eta$.

\section{Effect of finite time}
In this section, we investigate the effect of the finite time on the computation 
of the instantaneous frequencies.  We consider the signal 
$f(t)=e^{i\omega t}$ on $[0,T]$. 
\subsection{Windowed Fourier transform}
\begin{figure}
\centerline{\includegraphics*[scale=0.5]{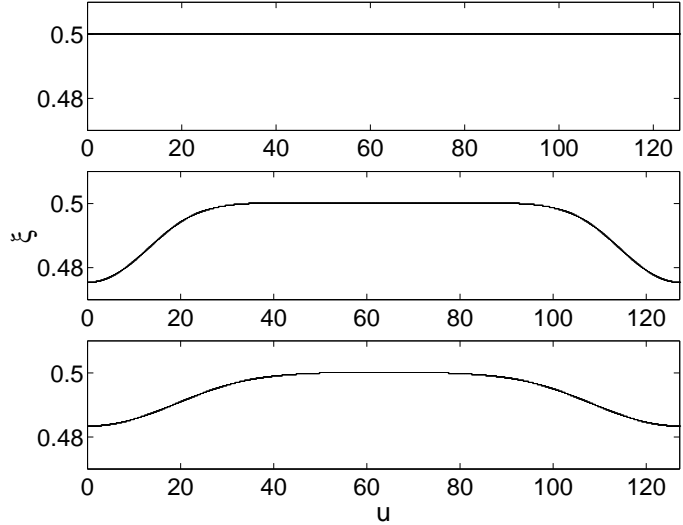}}
 \caption{\label{fig16} Wavelet ridge of a function $f(t)=e^{i \omega t}$ with 
$\omega=1/2$ on an interval $[0,T]$ for $(a)$ $\omega T=20\pi$, $(b)$ $\omega 
T=20\pi+\pi/4$ and $\sigma=1$, and $(c)$ $\omega T = 20\pi+\pi/4$ and 
$\sigma=1.5$.} \end{figure}

\begin{figure}
\centerline{\includegraphics*[scale=0.5]{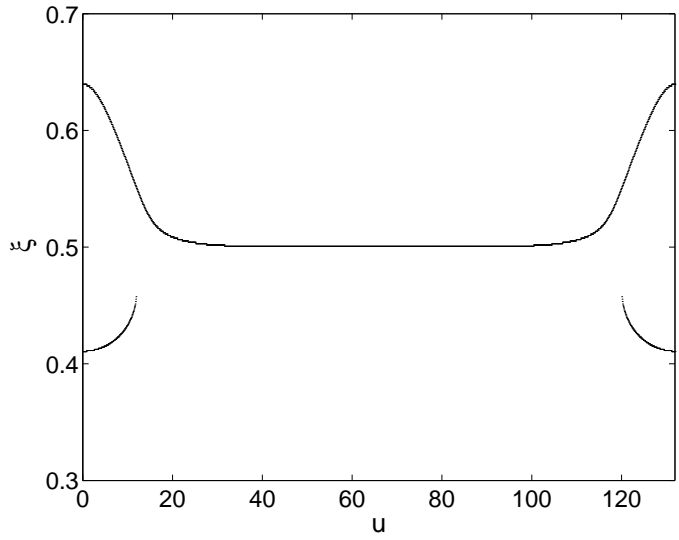}}
 \caption{\label{fig17} Wavelet ridge of a function $f(t)=e^{i \omega t}$ with 
$\omega=1/2$ on an interval $[0,T]$ for $\omega T=20\pi+\pi$.} \end{figure}

The windowed Fourier transform is computed by considering a periodic signal 
$\tilde{f}$ with period $T$ constructed from $f$, i.e.\ $\tilde{f}(t)=f(t)$ for 
$t\in [0,T[$ and $\tilde{f}(t+T)=\tilde{f}(t)$ for all $t$. However the signal 
$\tilde{f}$ may not be continuous if the value of the original signal $f$ at 0 
and $T$ are not the same. The discontinuities lead to an alteration of the 
instantaneous frequencies near these points.\\ We compute the ridge near a point 
of discontinuity, e.g., $u=0$. The windowed Fourier transform of $\tilde{f}$ is 
equal to $$
S\tilde{f}(u,\xi)=(\sigma^2\pi)^{-1/4} \sum_{n=-\infty}^{+\infty} 
\int_{nT}^{(n+1)T} \tilde{f}(t) e^{-i\xi t} e^{-(t-u)^2/(2\sigma^2)} dt .
$$
For $u$ close to zero, two terms are predominant in the sum: $n=-1$ and $n=0$. 
On the interval $[-T, 0]$, the function $\tilde{f}$ is equal to 
$e^{i\omega(t+T)}$. Thus
$$
(\sigma^2\pi)^{1/4} S\tilde{f}(u,\xi)\approx \int_0^T e^{i(\omega-\xi)t} 
e^{-(t-u)^2/(2\sigma^2)} dt +e^{i\omega T} \int_0^T e^{-i(\omega-\xi)t} 
e^{-(t+u)^2/(2\sigma^2)}. $$ 
We introduce the error function $\mbox{erf}(z)=\frac{2}{\sqrt{\pi}}\int_0^z 
e^{-t^2} dt$:
\begin{eqnarray*}
S\tilde{f}(u,\xi)=&&\left(\frac{\sigma^2\pi}{4}\right)^{1/4} 
e^{i(\omega-\xi)u}e^{-\sigma^2(\omega-\xi)^2/2} \\
&& \qquad \times \left[ 
\mbox{erf}\left(\frac{T-u-i\sigma^2(\omega-\xi)}{\sigma\sqrt{2}}\right)
+\mbox{erf}\left(\frac{u+i\sigma^2(\omega-\xi)}{\sigma\sqrt{2}}\right)
\right]                   \\
&&+e^{i\omega T }      \left(\frac{\sigma^2\pi}{4}\right)^{1/4} 
e^{i(\omega-\xi)u}e^{-\sigma^2(\omega-\xi)^2/2} \\
&& \qquad \times \left[ 
\mbox{erf}\left(\frac{T+u+i\sigma^2(\omega-\xi)}{\sigma\sqrt{2}}\right)
+\mbox{erf}\left(\frac{-u-i\sigma^2(\omega-\xi)}{\sigma\sqrt{2}}\right)
\right]
\end{eqnarray*}
We assume that $T$ is large compared to $u$ and $\sigma^2|\omega-\xi|$. 
We introduce the complex error function $w(z)$ as 
$w(z)=e^{-z^2}[1+\mbox{erf }(iz)]$ (see Ref.~\cite{abra64}). Since $\mbox{erf }z 
\to 1$ when $z\to \infty$ and $|\mbox{arg } z|<\pi/4$, the expression of the 
windowed Fourier transform becomes:
$$
S\tilde{f}(u,\xi)\approx  \left(\frac{\sigma^2\pi}{4}\right)^{1/4} 
e^{-y^2} \left[ 
w(x+i y) +e^{i\omega T} 
w(-x-i y)\right], $$ 
where $x=\sigma(\omega-\xi)/\sqrt{2} $ and $y=-u/(\sigma\sqrt{2})$.
We notice that when $\omega T=0 \mod 2\pi$, the expression of $S\tilde{f}$ 
becomes equal to
$$
S\tilde{f}(u,\xi)\approx    (4\pi\sigma^2)^{1/4} 
e^{iu(\omega-\xi)} e^{-\sigma^2(\omega-\xi)^2/2},
$$
since $w(z)+w(-z)=2e^{-z^2}$. That is the windowed Fourier transform of the 
periodic signal $e^{i\omega t}$ on $\mathbb{R}$. When $\omega T \not= 0 \mod 
2\pi$, the ridge is no longer constant and located at $\xi=\omega$, the 
discontinuity distorts locally the ridge. \\ We first consider $u=0$ for 
simplicity. Using the following development of the error function,
$$
w(z)=1+\frac{2i}{\sqrt{\pi}}z -z^2   -     
\frac{4i}{3\sqrt{\pi}}z^3+\frac{z^4}{2}+ \cdots , $$
we obtain the following expansion for the spectrogram:
\begin{eqnarray*}
P_S \tilde{f}(0,\xi)=&&2(1+\cos \omega T ) +x\frac{8}{\sqrt{\pi}}\sin \omega T 
-4x^2 \left[ 1-\frac{2}{\pi}+\left(1+\frac{2}{\pi}\right)\cos \omega T\right]\\ 
&& -x^3\frac{40}{3\sqrt{\pi}}\sin\omega T  
+4x^4  \left[ 1-\frac{8}{3\pi}+\left(1+\frac{8}{3\pi}\right)\cos \omega 
T\right] +\cdots ,\end{eqnarray*}                                                                       
where we have dropped the overall constant $\sigma\sqrt{\pi}/2$. 
The equation that determines the extrema of the spectrogram at $u=0$ is the 
following one:
\begin{eqnarray}
\frac{1}{\sqrt{\pi}}\sin \omega T 
&-& x \left[ 1-\frac{2}{\pi}+\left(1+\frac{2}{\pi}\right)\cos \omega T\right] 
-x^2\frac{5}{\sqrt{\pi}}\sin\omega T  \nonumber \\ &+& 2 x^3 \left[ 
1-\frac{8}{3\pi}+\left(1+\frac{8}{3\pi}\right)\cos \omega T\right] +\cdots  =0  
\label{eqn:solx0} \end{eqnarray}
From a solution $x_0$ of this equation, we obtain an 
approximation of the ridge:  $$
 \xi=\omega -\frac{\sqrt{2}}{\sigma} x_0 .
 $$
 One of the main features of a solution $x_0$ of this equation is that it does 
not depend on how large is the time interval. It only depends on the 
fractional part of $\omega T/(2\pi)$. Moreover this error decreases like 
$1/\sigma$ as $\sigma$ is increased. Again we notice that when $\omega T 
=0\mod 2\pi$ the ridge is located at $\xi=\omega$. \\
 We neglect the quadratic and cubic term of Eq.~(\ref{eqn:solx0}) which is 
valid if $\theta \equiv \omega T \mod 2\pi$ is close to zero. There is only one 
ridge at $u=0$ which is approximately located at
$$ \xi=\omega -\frac{1}{\sigma\sqrt{2\pi}}\sin \omega T. $$ The ridge is above 
$\xi=\omega$ when $\theta$ is negative, and below $\xi=\omega$ when $\theta$ is 
positive.\\ When $\omega T=\pi \mod 2\pi$, there is an extremum at $x_0=0$. 
However this solution is a minimum of the spectrogram. By considering the 
quartic term in the equation, we see that there are two other local minimum 
given by: $$ \xi_\pm=\omega \pm \frac{\sqrt{3}}{2\sigma}.
$$
Both ridges are local maxima of the spectrogram.\\
When the quadratic term of the 
expansion of the spectrogram vanishes (for $\omega T \approx 1.7947 \mod 
2\pi$), Eq.~(\ref{eqn:solx0}) shows that there are two extrema for the 
spectrogram: $$ \xi_{\pm}=\omega \pm\frac{1}{\sigma}\sqrt{\frac{5}{2}}.
$$
Only $\xi_-$ is a local maximum of the spectrogram and is located below 
$\xi=\omega$.\\ 
We compute next an approximation of the ridge when $u$ is 
increased. Using the expansion of $w$, we obtain \begin{eqnarray*}
P_S \tilde{f}(u,\xi)=&&P_S\tilde{f}(0,\xi) 
- 4x^2y^2 \left[ 1+\frac{2}{\pi}+\left(1-\frac{2}{\pi}\right)\cos \omega 
T\right]\\ &&  +x^3y^2\frac{40}{3\sqrt{\pi}}\sin\omega T 
 +4x^4y^2  \left[ 1-\frac{8}{3\pi}+\left(1+\frac{8}{3\pi}\right)\cos 
\omega T\right] \\ && +O(y^4,x^5) ,\end{eqnarray*}  
The equation that determines the ridge is the following one: 
\begin{eqnarray*}
&&\frac{1}{\sqrt{\pi}}\sin \omega T 
-x \left[ 1-\frac{2}{\pi}+\left(1+\frac{2}{\pi}\right)\cos \omega 
T\right]    -xy^2 \left[ 1+\frac{2}{\pi}+\left(1-\frac{2}{\pi}\right)\cos \omega 
T\right] \\
&&-x^2(1-y^2)\frac{5}{\sqrt{\pi}}\sin\omega T +2 x^3(1+y^2) \left[ 
1-\frac{8}{3\pi}+\left(1+\frac{8}{3\pi}\right)\cos \omega T\right] +\cdots  =0  
\end{eqnarray*}
The approximation of the ridge is given by
$$
\xi (u)=\omega -\frac{1}{\sigma\sqrt{2\pi}}
\left(1-\frac{u^2}{2\sigma^2}\right)\sin \omega T. $$
The ridge decreases (or increases) with $u$ if the ridge is above (resp.\ below) 
the main frequency $\omega$. The time interval where the effect of the 
discontinuity at $u=0$ is visible is approximately $[0 , \tau]$ where
$$
\tau =\sigma\sqrt{2}.
$$
The length of this interval is independent of the total time length $T$ and is 
proportional to $\sigma$. If we increase $\sigma$, the difference between the 
ridge and the frequency $\omega$ decreases as $1/\sigma$ but the time interval 
where this effect is visible increases as $\sigma$.  Figure~\ref{fig16} displays 
the ridge plots in three cases: The first panel is for $\theta=0$ where a 
flat ridge is expected, the second panel is for $\theta=\pi/4$ and 
$\sigma=1$, and the third panel is for $\theta=\pi/4$  and $\sigma=1.5$.\\ For 
$\omega T=\pi \mod 2\pi$, the extremum $x=0$ is a minimum for $y$ small and 
becomes a maximum for $y>1$. Thus the typical time of the effect of the 
discontinuity is again $\tau=\sigma\sqrt{2}$. For $y<1$, there are two ridges 
approximately located at :   $$ \xi_\pm (u)=\omega \pm 
\frac{\sqrt{3}}{2\sigma}\left(1-\frac{u^2}{2\sigma^2}\right). $$  These two 
ridges are symmetric with respect to $\xi(u)=\omega$. Figure~\ref{fig17} 
displays the two ridges in that case. The asymmetry on this figure is due to the 
correction of higher order terms in $x$ in the expansion of the spectrogram.

\subsection{Wavelet transform}
\begin{figure}   
\unitlength 1cm                                                               
\begin{picture}(10,10)
\put(0.4,0.7){\centerline{\includegraphics*[scale=0.5]{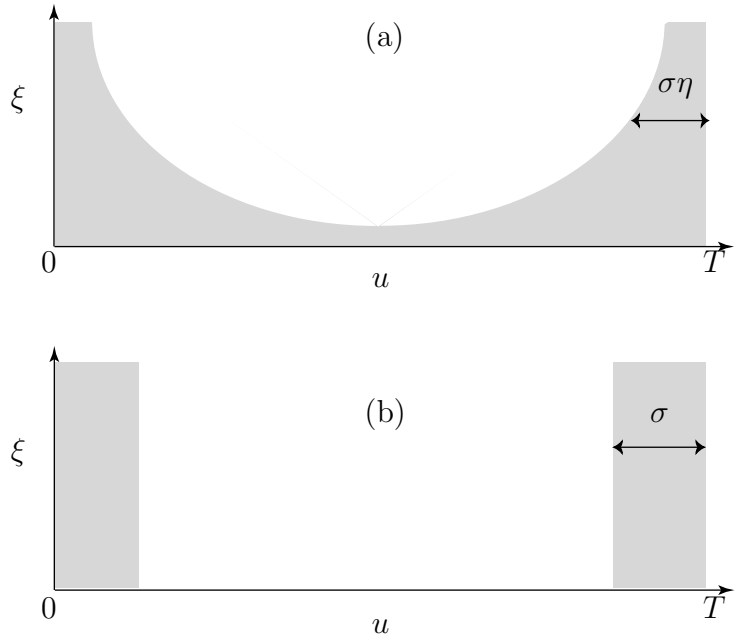}}  }
\put(2.7,0.4){0}
\put(7.1,0.2){$u$}
\put(11.5,0.4){$T$}
\put(2.7,5){0}
\put(7.1,4.8){$u$}
\put(11.5,5){$T$}
\put(7,3){(b)}
\put(7,8){(a)}
\put(2.3,2.5){$\xi$}
\put(2.3,7.2){$\xi$}
\put(10.8,3){$\sigma$}
\put(10.9,7.4){$\sigma\eta$}
\end{picture}
 \caption{\label{fig18} Schematic representation of the finite time effect 
on the computation of the ridges: The gray region is the part of the 
time-frequency plane where the discontinuity of the signal between $t=0$ and 
$t=T$ influences the computation of the instantaneous frequencies $(a)$ for 
the wavelet transform and $(b)$ for the windowed Fourier transform.} 
\end{figure}

The effect of a discontinuity at $u=0$ due to a periodisation is approximately 
the same as for the windowed Fourier transform. The computations are very 
similar. The wavelet transform of $\tilde{f}$ is given by
$$
W\tilde{f}(u,s)=(\sigma^2\pi)^{-1/4}
\frac{1}{\sqrt{s}} \sum_{n=-\infty}^{+\infty} 
\int_{nT}^{(n+1)T} \tilde{f}(t) e^{-i\eta (t-u)/s} e^{-(t-u)^2/(2s^2\sigma^2)} 
dt . $$
 Again two terms are important in the sum $n=-1$ and $n=0$. 
 \begin{eqnarray*}
e^{-iu\eta/s}\sqrt{s}(\sigma^2\pi)^{1/4} W\tilde{f}(u,s)\approx && \int_0^T 
e^{i(\omega-\eta/s)t} e^{-(t-u)^2/(2s^2\sigma^2)} dt \\ && +e^{i\omega T} 
\int_0^T e^{-i(\omega-\eta/s)t} e^{-(t+u)^2/(2s^2\sigma^2)}. 
\end{eqnarray*}
Next we assume that $T$ is large and we express the wavelet transform using the 
complex error function $w$:
$$
W\tilde{f}(u,s)\approx  \left(\frac{\sigma^2\pi}{4}\right)^{1/4} 
e^{-y^2} \left[ 
w(x+iy) +e^{i\omega T} w(-x-iy) \right], $$
where $x=\sigma(s\omega-\eta)/\sqrt{2}$ and $y=-u/(s\sigma\sqrt{2})$. 
The main difference 
with the windowed Fourier case is that now $y$ depends on the frequency (through 
the scale). For $\sigma\eta$ sufficiently large and $s$ sufficiently close 
to $\eta/\omega$, we get an approximation of $x$ and $y$:
\begin{eqnarray*}
&& x\approx \frac{\sigma\eta }{\omega\sqrt{2}} (\omega -\xi),\\
&& y\approx -\frac{\omega}{\sigma\eta\sqrt{2}} u,
\end{eqnarray*}
where $\xi=\eta/s$.
Thus the results of the windowed Fourier transform can be transposed for the 
wavelet transform by replacing $\sigma$ by $\sigma\eta/\omega$.  \\
Using the expansion of the error function, we obtain an approximation 
of the value of the maximum of the wavelet coefficients and hence an 
approximation of the ridge. For $u=0$ and for $\omega T \approx 0 \mod 2\pi$, 
this value is : $$
\xi=\omega -\frac{\omega}{\sigma\eta\sqrt{2\pi}}\sin \omega T.
$$
In the case of the wavelet ridge, we notice that the difference is proportional 
to the frequency, i.e.\ the the error in determining the frequency $\omega$ will 
be smaller for low frequencies than for large ones.\\
For $u$ close to zero, we get an approximation of the ridge:
$$
\xi(u)=\omega -\frac{\omega}{\sigma\eta\sqrt{2\pi}}
\left(1-\frac{u^2\omega^2}{2\sigma^2\eta^2}\right)\sin \omega T. $$
 The typical 
width in time of the error in determining the frequency $\omega$ is given by
$$
\tau\approx \frac{\sigma\eta\sqrt{2}}{\omega}.
$$
We notice that increasing the parameters of the wavelet decrease the amplitude 
of the error but increases the typical length time where this error is 
visible.\\
The wavelet method cannot determine frequencies such that  $\tau\geq T/2$, i.e.\ 
frequencies such that $$
\omega \leq 2\sqrt{2} \frac{\sigma\eta}{T}
$$
In summary, the fact that the trajectory is known on a finite time 
interval creates in general a discontinuity on the boundaries of this interval. 
This distorts the ridge on a small time interval. It creates one or two 
ridges; when there are two ridges, it means that the main original ridge is a 
minimum of the spectrogram. When there is one ridge, it can be below or above 
the main ridge depending on the value of the fractional part of $\omega 
T/(2\pi)$. We have seen that the amplitude of this distortion is inversely 
proportional to the size of the window of the Fourier transform and that the 
typical time interval during which this distortion is visible is proportional to 
this size of the window.   \\
Figure~\ref{fig18} represents the influence of the discontinuity in the 
time-frequency plane. The gray region is the portion of the time-frequency 
plane that has to be discarded due to the influence of the discontinuity on the 
boundaries. For the windowed Fourier transform, this error does not depend on 
the frequency whereas for the wavelet transfrom, the distorsion influences 
greatly the low frequencies.

\end{document}